\documentclass[aps,book,onecolumn,showpacs,preprint]{revtex4-1}
\usepackage{epsfig,epsf}
\pdfoutput=1
\usepackage{graphicx,tabularx}
\newcolumntype{Y}{>{\centering\arraybackslash}X}
\usepackage{epstopdf}
\usepackage{bm}
\usepackage{color,esvect}
\usepackage{hyperref}
\setcitestyle{authoryear,round}
\begin{document}

\title{Water between membranes: Structure and Dynamics.}

\author{Sotiris Samatas}
\altaffiliation{%
Secci\`o de Fisica Estadistica i Interdisciplinaria--
Departament de F\'isica de la Mat\`eria Condensada
 \&      Institute of Nanoscience and Nanotechnology (IN2UB),
Universitat de Barcelona, Mart\'i Franqu\'es 1, 08028 Barcelona}

\author{Carles Calero}
\altaffiliation{%
Secci\`o de Fisica Estadistica i Interdisciplinaria--
Departament de F\'isica de la Mat\`eria Condensada
 \&      Institute of Nanoscience and Nanotechnology (IN2UB),
Universitat de Barcelona, Mart\'i Franqu\'es 1, 08028 Barcelona}

\author{Fausto Martelli}
\altaffiliation{%
IBM Research, Hartree Centre, Daresbury WA4 4AD, United Kingdom
}

\author{Giancarlo Franzese$^*$}
%\email{carles.calero@ub.edu}

\email{gfranzese@ub.edu}

\begin{abstract} %150 ? 200 words
An accurate description of the structure and dynamics of interfacial water is essential for phospholipid membranes, since it determines their function and their interaction with other molecules. Here we consider water confined in stacked membranes with hydration from poor to complete, as observed in a number of biological systems. Experiments  show that the dynamics of water slows down dramatically when the hydration level is reduced. 
All-atom molecular dynamics simulations identify three (inner, hydration and outer) regions, within a distance of approximately 1 nm from the membrane, where water molecules exhibit different degrees of slowing down in the dynamics.   
The slow-down is a consequence of the robustness of the hydrogen bonds between water and lipids  and the long lifetime of the hydrogen bonds between water molecules near the membrane. 
The interaction with the interface, therefore, induces a structural change in the water 
that can be emphasized by calculating its intermediate range order. Surprisingly,  at distances as far as $\simeq$ 2.5 nm from the interface, although the bulk-like dynamics is recovered, the intermediate range order of water is still slightly higher than that in the bulk at the same thermodynamic conditions. 
Therefore, the water-membrane interface has a structural effect at ambient conditions that propagates further than the often-invoked 1 nm length scale. 
Membrane fluctuations smear out this effect macroscopically, but an analysis performed by considering local distances and instantaneous configurations is able to reveal it, possibly contributing to our understanding of the role of water at biomembrane interfaces.
\end{abstract}


\maketitle

\section{Introduction}

It has been long recognized that the structure and function of biological membranes is greatly determined by the properties of hydration water (Berkowitz et al. 2006).  Biological membranes are made of a large number of components, including membrane proteins, cholesterol, glycolipids, ionic channels among others, but their framework is provided by phospholipid molecules forming a bilayer. The bilayer structure is a consequence of the interaction with water of the phospholipids, that  are macromolecules made of a hydrophilic end (headgroup) and a hydrophobic side (chain or tailgroup). To reduce the free-energy cost of the interface, the polar heads expose to water,  with their tails side by side, while the apolar hydrocarbon tails hide from water extending in the region between two layers of heads (hydrophobic effect). 

But water is important for membranes not only because it induces the hydrophobic effect. On the one hand, interfacial water modulates the fluidity of the membrane, on the other hand it mediates the interaction between membranes and between membranes and solutes (ions, proteins, DNA, etc.), regulating cell-membrane tasks as, for example, transport and signaling functions (Hamley 2007).  

Phospholipid bilayers or monolayers of a single type of phospholipid are used  as model systems to understand how basic biological membranes function and how they interact with the environment. For its important role in determining the  properties of biological membranes, structure and dynamics of water at the interface with phospholipid bilayers have been extensively investigated 
both in experiments and in simulation studies.

Experiments have used Nuclear Magnetic Resonance (NMR) spectroscopy  to study the translational dynamics of interfacial water, evidencing the different rates of lateral and normal diffusion and revealing the effect of lipid hydration on water dynamics (Volke et al. 1994, Wassall 1996). 
The slow-down of water dynamics due to the interaction with the phospholipid membrane has also been observed with the help of molecular dynamics (MD) simulations (Berkowitz et al. 2006, Bhide and Berkowitz 2005). 

NMR experiments and vibrational sum frequency generation spectroscopy have provided insight on the ordering and orientation of water molecules around phospholipid headgroups (Chen et al. 2010, K\"{o}nig et al. 1994). These results are in agreement with the picture extracted from computer simulations of hydrated phospholipid membranes (Berkowitz et al. 2006, Pastor 1994).  

Infrared spectroscopy measurements indicate the formation of strong hydrogen bonds with the phosphate and carbonyl groups of phospholipids, as well as an enhancement of the hydrogen bonds between water molecules in the vicinity of phospholipid headgroups (Binder 2003, Chen et al. 2010). The rotational dynamics of water molecules is also dramatically affected by the presence of phospholipids and the hydration level of the membrane, as evidenced experimentally using a variety of techniques including ultrafast vibrational spectroscopy (Zhao et al. 2008), terahertz spectroscopy (Tielrooij et al. 2009), and neutron scattering (Trapp et al. 2010). MD simulations have complemented these studies by exploring the decay of water orientation correlation functions in phospholipid membranes with different hydration levels (Zhang and Berkowitz 2009, Gruenbaum and Skinner 2011, Calero et al. 2016, Martelli, Ko, Borallo and Franzese 2018).

These studies reveal that the structure and dynamics of hydration water in stacked phospholipid membranes are affected by both the interaction with phospholipids and by its level of hydration. In the following, we present the results of studies based on all-atom MD simulations which probe the structural and dynamical properties (both diffusion and rotational dynamics) of hydration water at stacked phospholipid membranes as a function of their hydration. 

First, the focus is put on the influence of the membrane hydration level on the translational and rotational dynamical properties of confined water. Second, we perform a detailed analysis to determine the dependence of water dynamics on the local distance to the membrane and on temperature. We also relate the dynamical behavior of water confined between bilayers with the structure and dynamics of the hydrogen bond network formed with other water molecules and with selected groups of the lipid.  

We finally discuss the surprising result showing that the water-membrane interface has a structural effect at ambient conditions that propagates further than the often-invoked 1 nm length scale. To this goal we discuss the calculation of water intermediate range order at the interface adopting a parameter recently introduce  by Martelli et al. (Martelli, Ko, O\u{g}uz and Car 2018).

\begin{figure}%[H]
\begin{center}
\vspace*{0.5cm}
\includegraphics*[angle=0, width=10cm]{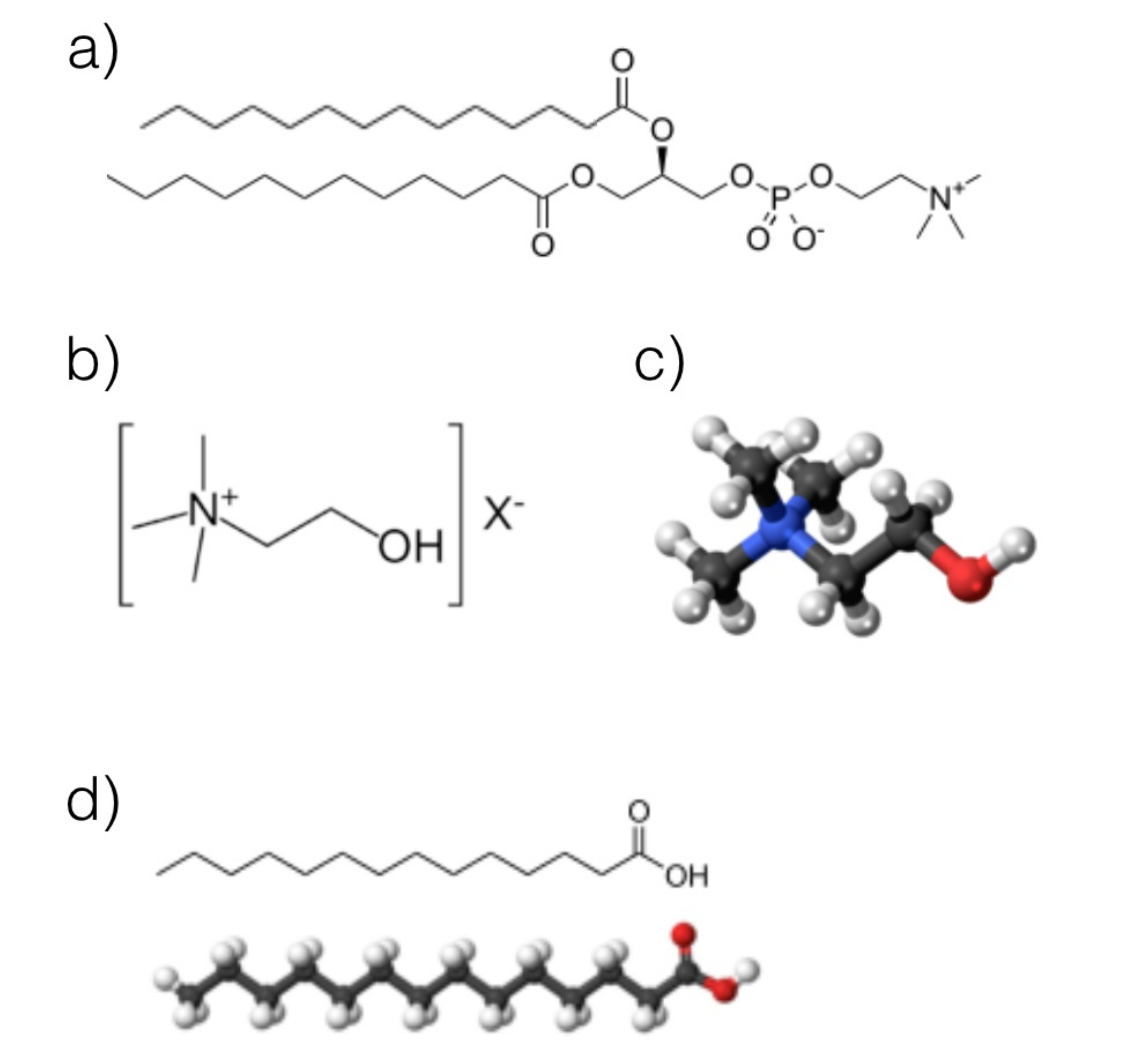}
\caption{Dimyristoylphosphatidylcholine (DMPC) phospholipid macromolecule. 
a) DMPC chemical structure. 
b) Headgroup: choline (quaternary ammonium salt) containing the N,N,N-trimethylethanolammonium cation and with an undefined counteranion X$^-$.
c)  Choline detailed structure as ball-and-stick, with carbon in black, hydrogen in white, oxygen in red, nitrogen in blue.
d)  Tailgroup: two myristoyl chains,  in skeletal and ball-and-stick representations.} 
\label{Fig:DMPC}
\end{center}
\end{figure}

\section{Dynamics of water  between stacked membranes at different hydration levels}\label{overalldynamics}

In the following we focus on  water  between stacked membranes,  
an important part in several biological structures, including endoplasmic reticulum and Golgi apparatus, that processes proteins for their use in animal cells,  or thylakoid compartments in chloroplasts (plant and algal cells) and cyanobacteria, both  involved in the photosynthesis process. We will consider the water dynamics when the membrane hydration level changes, and will use as a model membrane made of 
dimyristoylphosphatidylcholine (DMPC) phospholipid bilayers. 

Among a wide variety of lipids, DMPC are phospholipids incorporating a choline as a headgroup and a tailgroup formed by two myristoyl chains (Figure~\ref{Fig:DMPC}). Choline based phospholipids are ubiquitous in cell membranes and used in drug targeting liposomes (Hamley 2007). 

\begin{figure}%[H]
\begin{center}
\vspace*{0.5cm}
\includegraphics*[angle=0, width=16cm]{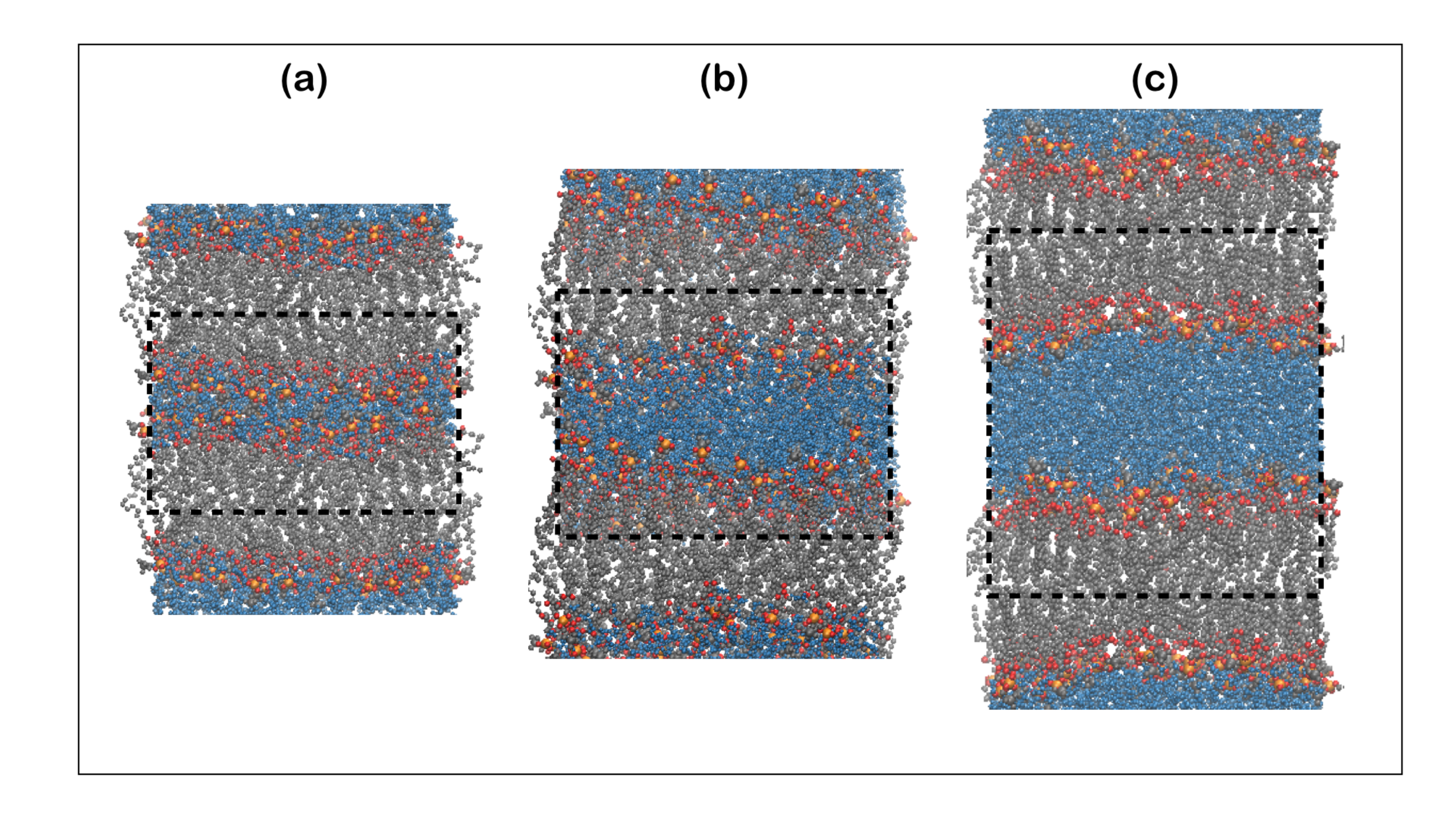}
\caption{Snapshots of a DMPC bilayer with fixed hydration levels (a) $\omega =$ 7, (b) 20, (c) 34.   
Other hydration levels considered in (Calero et al 2016) were $\omega =$ 4, 10, and 15.  Gray and red beads represent phospholipid tails and headgroups,
  respectively. Blue and white beads represent oxygen and hydrogen
  atoms of water. The dashed line indicates the size of the simulation
  box. } 
\label{Fig:snapshots}
\end{center}
\end{figure}

In MD simulations (Calero et al. 2016),  by using periodic boundary conditions  (Figure~\ref{Fig:snapshots}),  it is possible to describe a system of perfectly stacked phospholipid bilayers with a homogeneous prescribed hydration level $\omega$, including the low hydrated systems probed in experiment (Tielrooij et al. 2009, Trapp et al. 2010, Volkov et al. 2007, Zhao et al. 2008) and a fully hydrated membrane (with $\omega = 34$) (Nagle et al.1996). The last case has been thoroughly studied  using computer simulations (Bhide and Berkowitz 2005). 

In particular,
Calero et al. (Calero et al. 2016) consider 128 DMPC lipids distributed in two
leaflets in contact with hydration water, and perform MD simulations using the NAMD 2.9
(Phillips et al. 2005) package at a temperature of 303 K and an
average pressure of 1 atm, setting the simulation time step to 1 fs, and  
describing the structure of phospholipids and their mutual interactions
by the recently parameterized force field CHARMM36
(Klauda  et al 2010, Lim  et al 2012), which is able to
reproduce the area per lipid in excellent agreement with experimental
data. The water model employed in their simulations, consistent with the
parametrization of CHARMM36, is the modified TIP3P
(Jorgensen  et al 1983, MacKerell  et al 1998).  They cut off the van
der Waals interactions at 12\AA\ with a smooth switching function
starting at 10\AA\ and compute the long-range electrostatic forces using
the particle mesh Ewald method (Essmann  et al 1995) with a grid space
of $\approx 1$\AA, updating the electrostatic interactions every
2~fs. After energy minimization, they equilibrate the hydrated
phospholipid bilayers for 10~ns followed by a production run of 50~ns in
the NPT ensemble at 1 atm. They use a Langevin
thermostat (Berendsen  et al 1984) with a damping coefficient of
0.1~ps$^{-1}$ to control the temperature and a Nos\'{e}-Hoover
Langevin barostat (Feller  et al 1995) with a piston oscillation
time of 200~fs and a damping time of 100~fs to control the pressure.

\subsection{Translational dynamics}

\begin{figure}%[H]
\begin{center}
\vspace*{0.5cm}
\includegraphics*[angle=0, width=11cm]{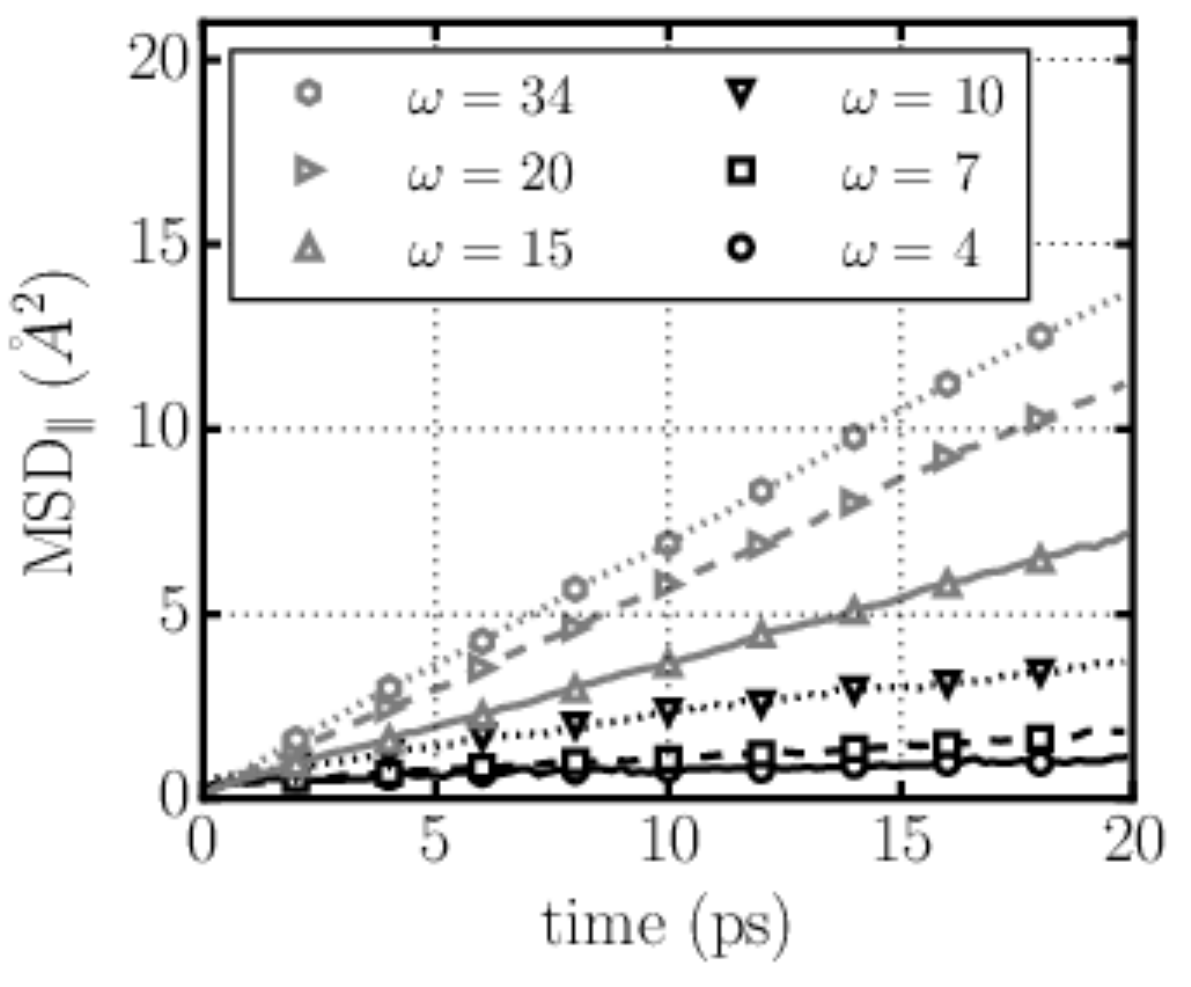}
\includegraphics*[angle=0, width=11cm]{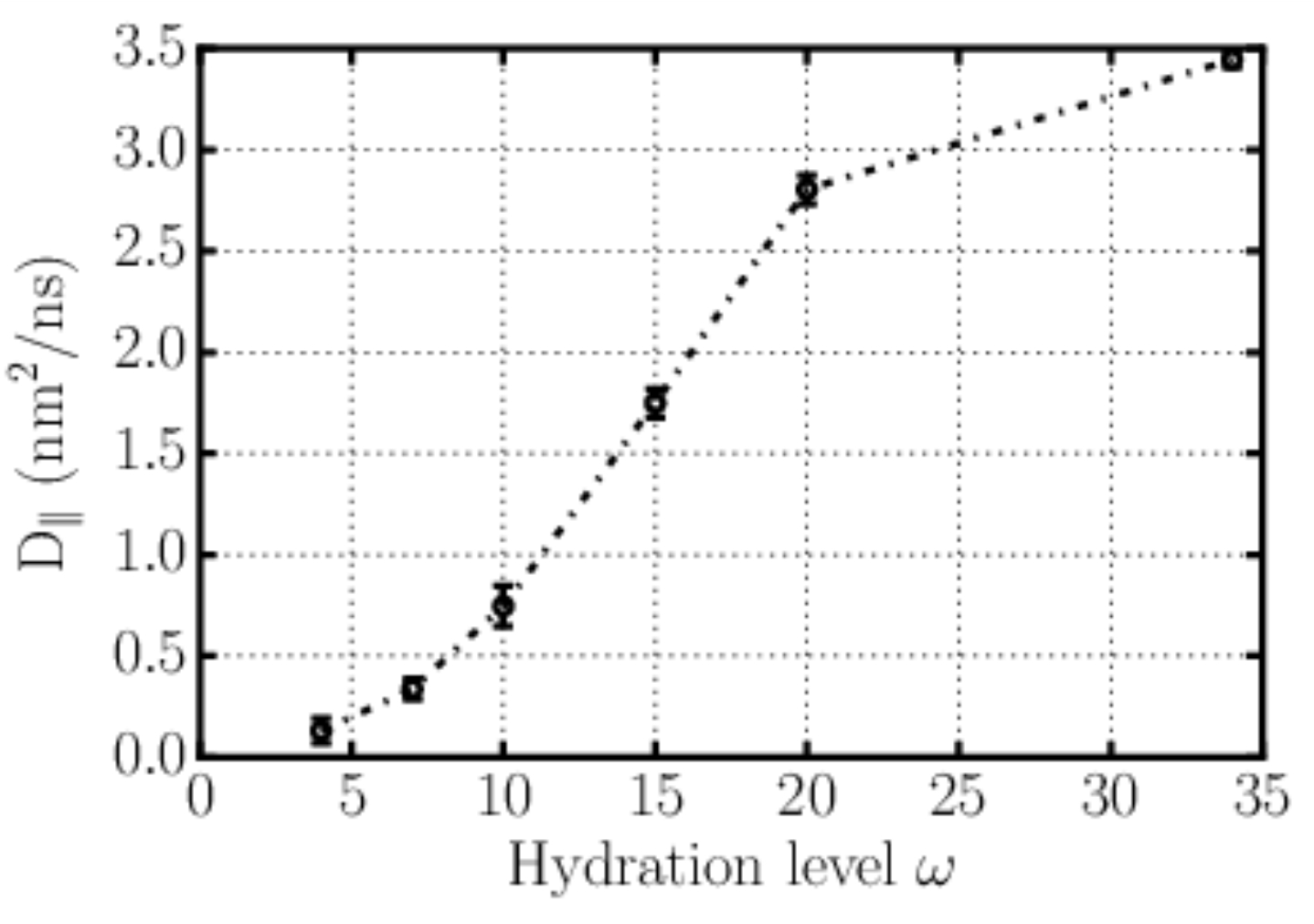}
\caption{Translational dynamics of confined water molecules projected on the
  plane of the membrane for the different stacked phospholipid
  bilayers. 
  (Upper panel) Mean-square displacement MSD$_{\parallel}$ on the plane of the membrane, as a function of time. 
  (Bottom panel) Diffusion coefficient $D_{\parallel}$ 
  of water molecules on the plane of the membrane for the different
  hydration levels considered. Adapted from Adapted from (Calero et al. 2016).} 
\label{Fig:diffcoeff}
\end{center}
\end{figure}

The translational dynamics of water confined in stacked DMPC bilayers is characterized by the mean-square displacement of the center of mass of water molecules projected onto the plane of the membrane (MSD$_{\parallel}$). The diffusion coefficients for confined water are obtained from the linear regime reached by the MSD at sufficiently long times through
\begin{equation}
D_{\parallel} \equiv \lim_{t \to \infty} \frac{\langle |{\bf r}_{\parallel}(t) - {\bf r}_{\parallel}(0)|^2   \rangle}{4t}\,,
\end{equation}
where ${\bf r}_{\parallel}(t)$ is the projection of the center of mass of a water molecule on the plane of the membrane and the angular brackets $\langle ... \rangle$ indicate average over all water molecules and time origins (Figure~\ref{Fig:diffcoeff}).

Calero et al. (Calero et al 2016) find that water molecules are significantly slowed down when the hydration level of the membrane is reduced, in agreement with previous experimental and computational studies (Wassall 1996, Zhao  et al 2008, Tielrooij  et al 2009, Zhang and Berkowitz 2009, Gruenbaum and Skinner 2011). The diffusion coefficient increases monotonically with hydration (Figure~\ref{Fig:diffcoeff}), from $D_{\parallel} = 0.13$ nm$^2$/ns for the lowest hydrated system ($\omega= 4$) to 3.4 nm$^2$/ns for the completely hydrated membrane ($\omega= 34$), in  agreement with experimental results of similar systems (Wassall 1996, Rudakova  et al 2004). 
The agreement is not quantitative, because a  comparison of the MD results  with experimental results is problematic due to the difficulty in experiment to ensure the homogeneity of hydration and the perfect alignment of the membranes in measuring $D_{\parallel}$.  
Nevertheless, the MD simulations reproduce  qualitatively the large drop of the diffusion coefficient $D_{\parallel}$ for low hydrated membranes with respect to bulk water and its dependence with hydration, as seen 
in (Wassall 1996), where  the diffusion of water confined in the lamellar phase of egg
phosphatidylcholine was investigated as a function of the hydration of the membrane. For weakly hydrated systems the authors of the experiments observed that the water diffusion coefficient had an important reduction (of approximately a factor 10), which was attributed to the interaction with the membrane. In addition, the authors found a monotonic increase of the diffusion coefficient with the membrane's hydration. 
It is, therefore, clear that the MD approach is able, at least qualitatively, to reproduce the translational behavior of water between stacked membranes, suggesting the possibility to get insight into the detailed mechanisms of the  slowing down at the molecular level.

\subsection{Rotational dynamics}
\label{sec:tau}

\begin{figure}%[]
\begin{center}
\vspace*{0.5cm}
\includegraphics*[angle=0, width=11cm]{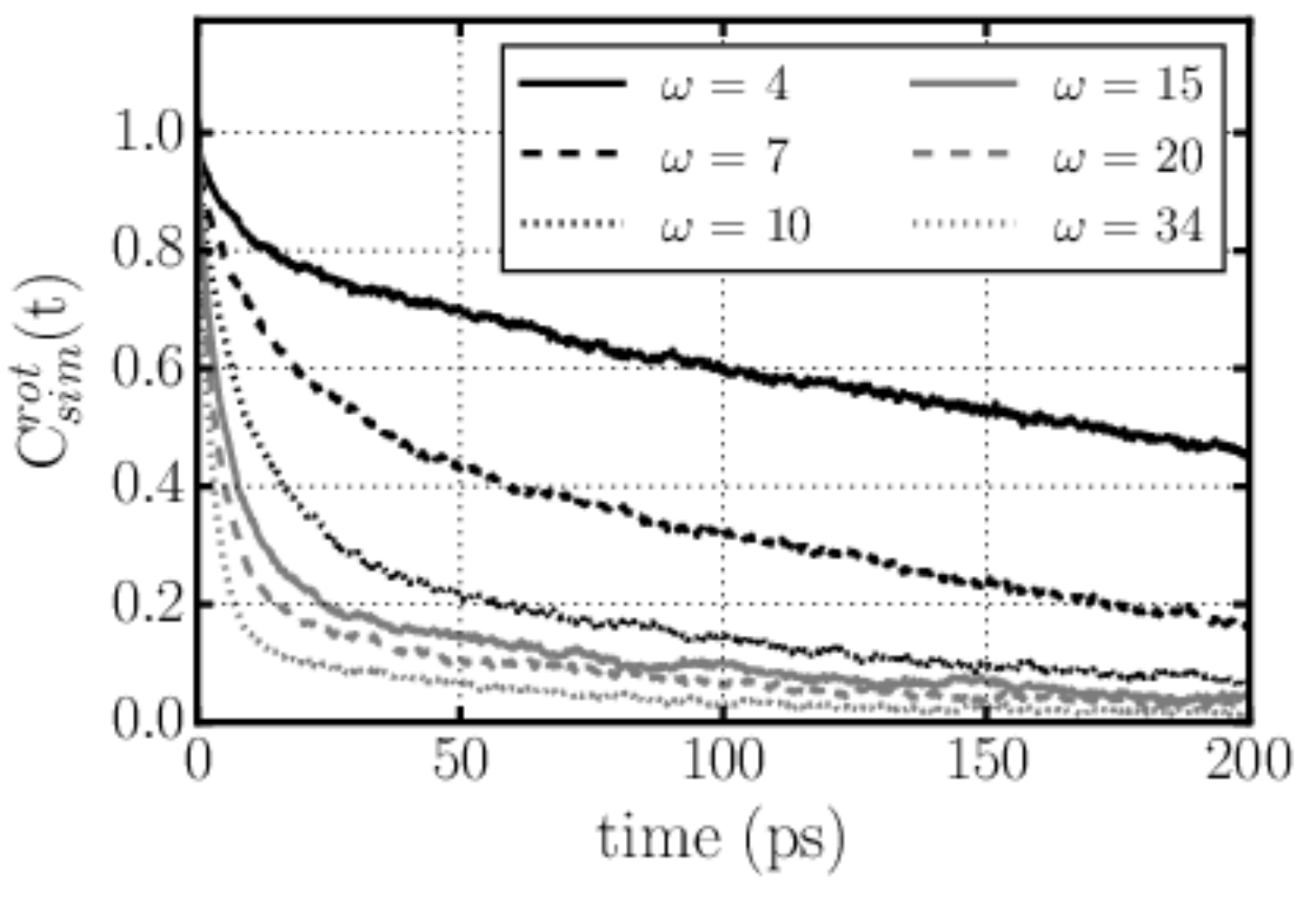}
\includegraphics*[angle=0, width=11cm]{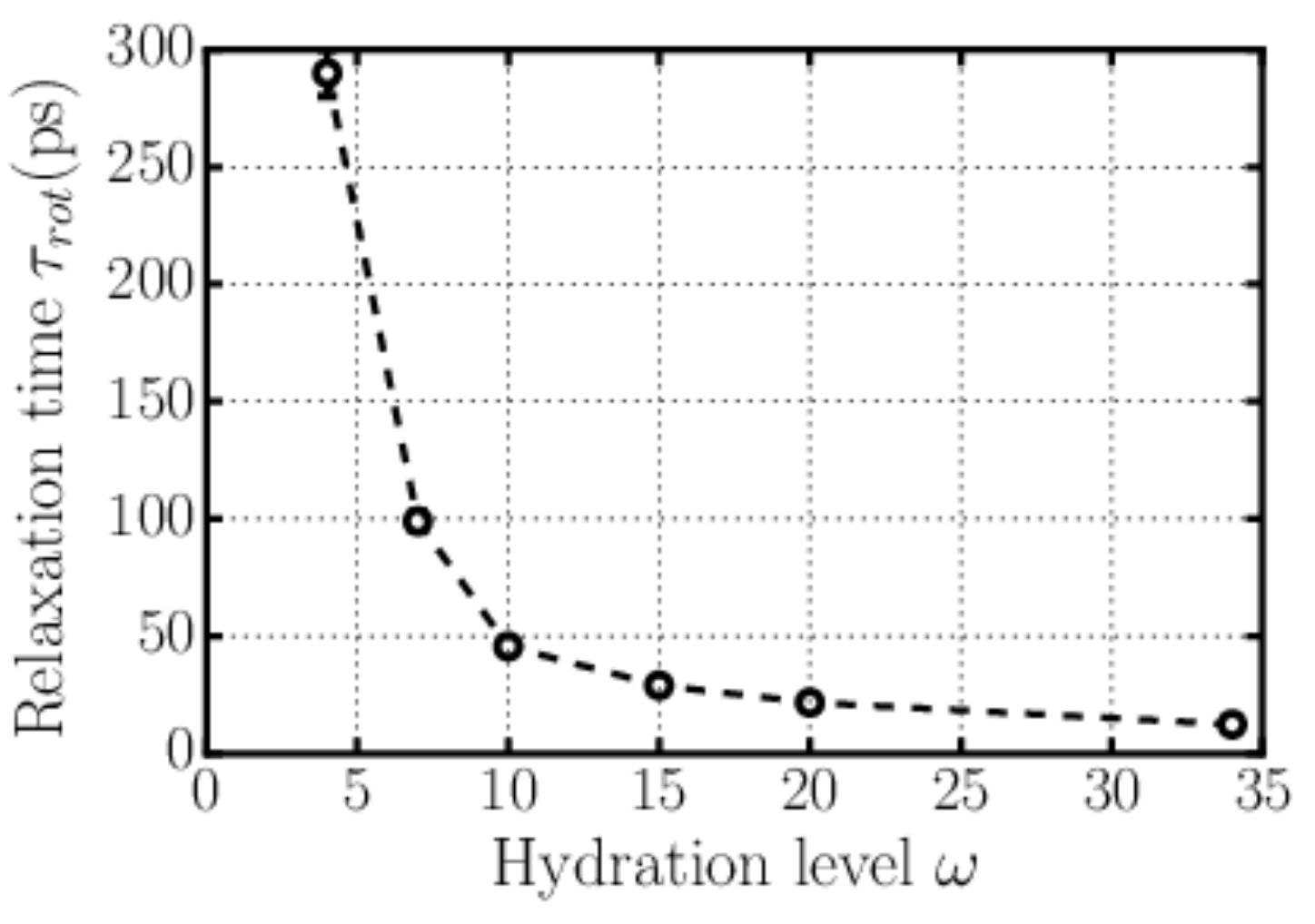}
\caption{Rotational dynamics of  water molecules confined in stacked phospholipid bilayers at different levels of hydration $\omega$.  
(Upper panel) 
Rotational dipolar correlation function $C_{\rm sim}^{\rm rot}$ of water
molecules for $\omega=$ 4, 7, 10, 15, 20, 34 (from top to bottom).  
(Lower panel) Rotational relaxation time $\tau_{\rm rot}$ calculated from Eq.(\ref{tau}) as a function of the hydration level. Adapted from (Calero  et al 2016).} 
\label{Fig:rotcorrfunctiona}
\end{center}
\end{figure}

The rotational dynamics of the water molecules confined in stacked phospholipid bilayers is characterized by calculating from the trajectories of the MD simulations the rotational dipolar correlation function,  
\begin{equation}\label{Eq:Crot}
 C_{\rm sim}^{\rm rot}(t) \equiv \langle \hat{\mu}(t)\cdot \hat{\mu}(0) \rangle\,,
\end{equation}
where $\hat{\mu}(t)$ is the direction of the water dipole vector at time $t$ and $\langle ... \rangle$ denote ensemble average over all water molecules and time origins (Figure~\ref{Fig:rotcorrfunctiona}). This quantity is related to recent terahertz dielectric relaxation measurements used to probe the reorientation
dynamics of water (Tielrooij et al 2009). 

To quantify the relaxation of the correlation functions $C_{\rm sim}^{rot}(t)$ Calero et al. (Calero  et al 2016)  define the relaxation time 
\begin{equation}
 \tau_{\rm rot} \equiv \int_0^{\infty} C_{\rm sim}^{\rm rot}(t) dt\,,
\label{tau}
\end{equation}
which is independent of any assumptions on the functional form of the correlation function. Calero et al. find that $\tau_{\rm rot}$ decreases from (290 $\pm$ 10) ps at low hydration $\omega$=4, to  (12.4 $\pm$ 0.3) ps at full hydration $\omega$=34 (Figure~\ref{Fig:rotcorrfunctiona}), with
a monotonic slowing-down of the rotational dynamics of water for decreasing membrane hydration. 
This result is consistent with  experiments (Zhao et al 2008, Tielrooij et al. 2009) and  previous computational works (Zhang and Berkowitz 2009, Gruenbaum and Skinner 2011).

\begin{figure}%[]
\begin{center}
\vspace*{0.5cm}
\includegraphics*[angle=0, width=10cm]{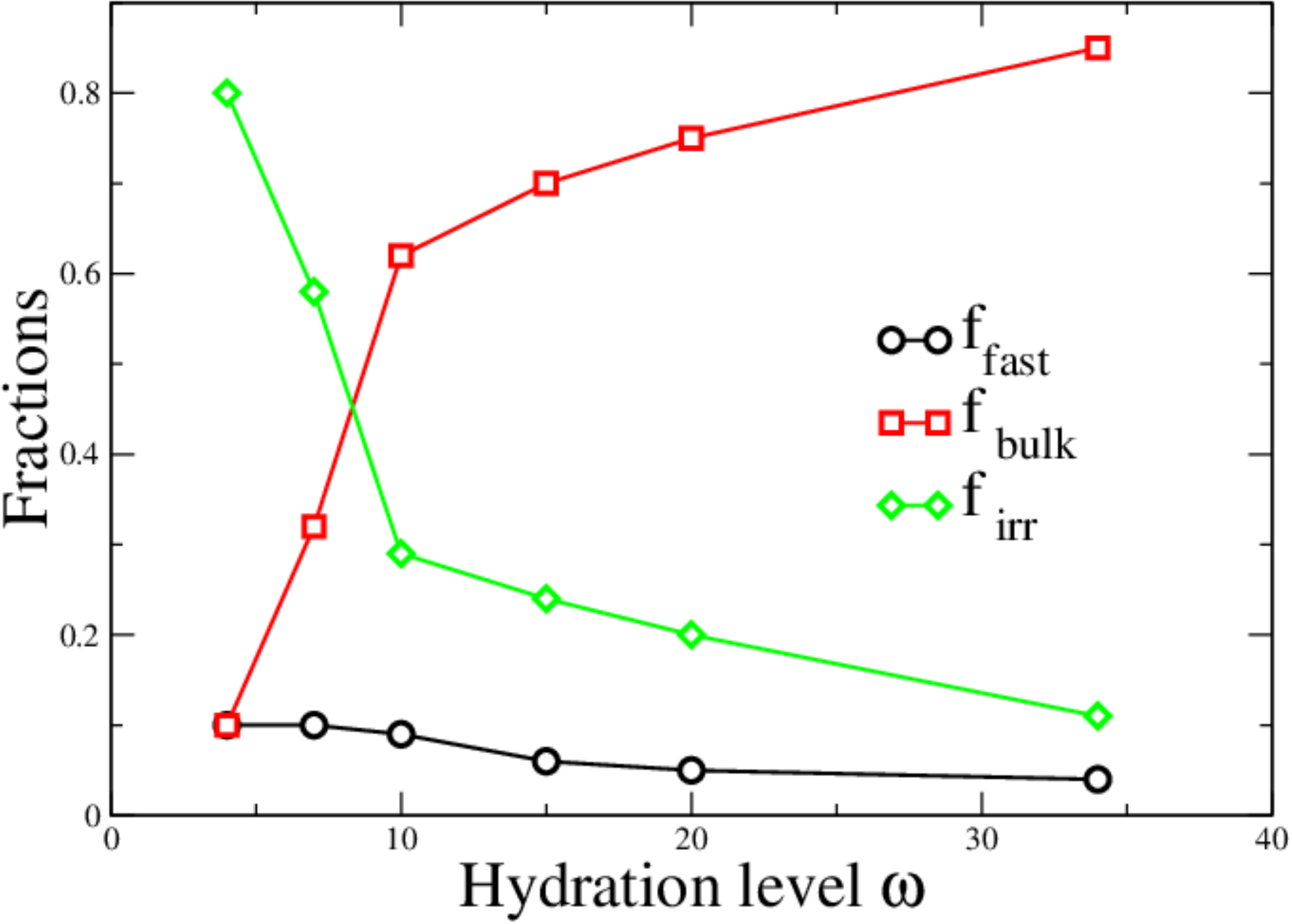}
\includegraphics*[angle=0, width=10cm]{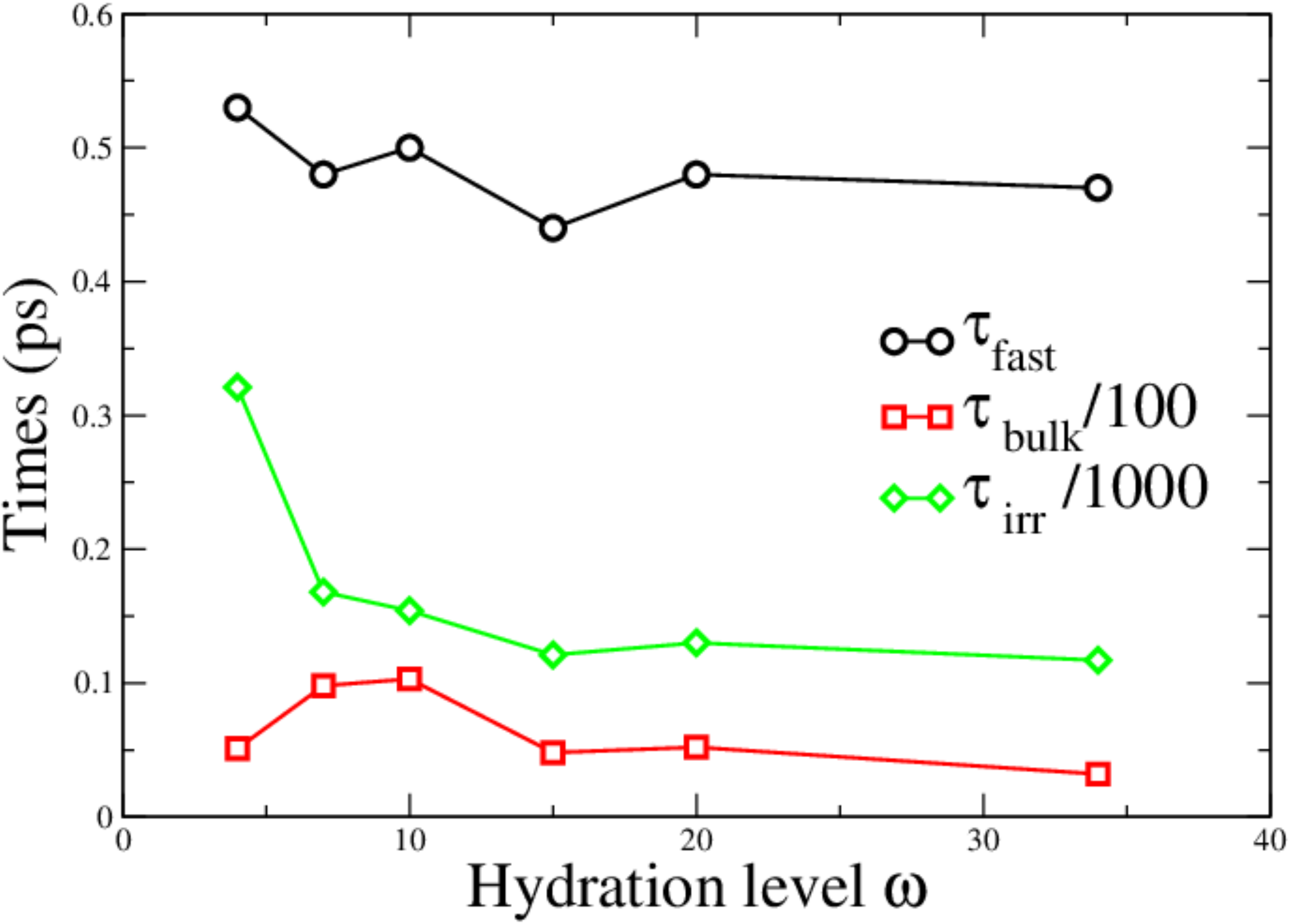}
\caption{Partition  into fast, bulk-like and  irrotational water molecules as in Eq.(\ref{C_fit}).
(Upper panel)  Fractions  $f_{\rm fast}$, $f_{\rm bulk}$, and $f_{\rm irr}$ with the condition in Eq.(\ref{condition}).
(Lower panel) Characteristic times for each of the three species, 
 $\tau_{\rm fast}$, $\tau_{\rm bulk}$, and $\tau_{\rm irr}$, respectively. For sake of clarity, times are rescaled as indicated in the legend. The characteristic times have no regular behavior  for increasing hydration.
 Adapted from (Calero et al. 2016). } 
\label{Fig:rotcorrfunctionb}
\end{center}
\end{figure}

The decay of the rotational correlation function $C_{\rm sim}^{\rm rot}(t)$ represented in Figure~\ref{Fig:rotcorrfunctiona} occurs through an initial rapid decrease of
$C_{\rm sim}^{\rm rot}(t)$ followed by a much slower decay process. This dependence suggests the existence of different time-scales relevant in the rotational dynamics of water molecules. In particular, Tielrooij et al. 
 assumed the existence of three different water species in regards to their reorientation dynamics   (Tielrooij et al. 2009): (i) {\it bulk-like water} molecules, whose reorientation dynamics resemble that of bulk water, with characteristic times $\tau_{\rm bulk}$ of a few picoseconds; (ii) {\it fast water} molecules, which reorient, significantly faster than bulk water with $\tau_{\rm fast} \approx$ fraction of picosecond, and (iii) {\it irrotational water} molecules, which might relax with characteristic times $\tau_{\rm irr} \gg 10$ ps. With this assumption, one can identify the fractions $f_{\rm fast}$, $f_{\rm bulk}$ and $f_{\rm irr}$ of the three species as a function of $\omega$ by fitting $C_{\rm sim}^{\rm rot}(t,\omega)$ to a sum of pure exponentially decaying terms: 
\begin{equation}
 C_{fit}(t,\omega) = f_{\rm fast}(\omega) e^{-t/\tau_{\rm fast}} + f_{\rm bulk}(\omega) e^{-t/\tau_{\rm bulk}} + f_{\rm irr}(\omega) e^{-t/\tau_{\rm irr}}\,,
\label{C_fit}
\end{equation}
with
\begin{equation}
  f_{\rm fast}(\omega) + f_{\rm bulk}(\omega) +  f_{\rm irr}(\omega) = 1
\label{condition}
\end{equation}
for each $\omega$.

Using this fitting procedure, Calero et al. (Calero et al. 2016)  analyze their MD results for water molecules for different hydration levels of the bilayer (Figure \ref{Fig:rotcorrfunctionb}). 
Their results  qualitatively account for the experimental behavior reported in Ref. (Tielrooij et al. 2009). However, such a fitting procedure is not robust--there are five fitting parameters for each $\omega$--and the parameters $\tau_{\rm fast}$,  $\tau_{\rm bulk}$ and $\tau_{\rm irr}$ do not show any regular behavior as function of $\omega$. This strongly suggests that the hypothesis made by Tielrooij et al.  of the existence of such distinctive types of water might not be complete and that a more thorough analysis is needed, as described in the next section.

\section{Dependence of water dynamics on distance to membrane}

\subsection{Water structure at stacked phospholipid bilayers}\label{structure}
\subsubsection{Definition of distance to the membrane}
\label{subIIIA1}

To understand the dependence of the dynamics of water molecules on membrane hydration, a detailed investigation of the water-membrane interface
 is necessary. To this end, a suitable definition of a distance to the membrane is required. In the relevant length-scales of the problem (defined, for example, by the size of the water molecule) the water-membrane interface is not flat, but exhibits spatial inhomogeneities of $\approx 1$nm. In addition, these inhomogeneities are dynamical since the phospholipid membrane is in a two dimensional fluid phase. Nevertheless, its dynamics is much slower than water dynamics, with characteristic timescales significantly longer than the relevant timescales of water dynamics, as evidenced by the disparity of diffusion coefficients: while the typical diffusion coefficient of water is of the order of 1 nm$^2$/ns, the diffusion coefficient of phospholipids is of $\approx 0.001$ nm$^2$/ns (Yang et al. 2014). Additionally, the interface is soft and water molecules can penetrate into the membrane (Fitter et al. 1999, Pandit et al. 2003, Lopez et al. 2004). 

In order to describe the interface and accomodate such features of the membrane, Pandit, Bostick, and Berkowitz (Pandit et al. 2003, Berkowitz et al. 2006) devised a local definition of the distance to the membrane. Briefly, for each frame a 2-dimensional Voronoi tessellation of the plane of the membrane (the XY-plane) is performed using as centers of the cells the phosphorous and nitrogen atoms of the phospholipid heads. Each water molecule is assigned a Voronoi cell by its location in the XY plane and has a distance,  $\xi$,  to the membrane given by the difference between the Z-coordinates of the water molecule and its corresponding Voronoi cell. 

\begin{figure}[h!]
\begin{center}
\vspace*{0.cm}
\includegraphics*[angle=0, width=11cm]{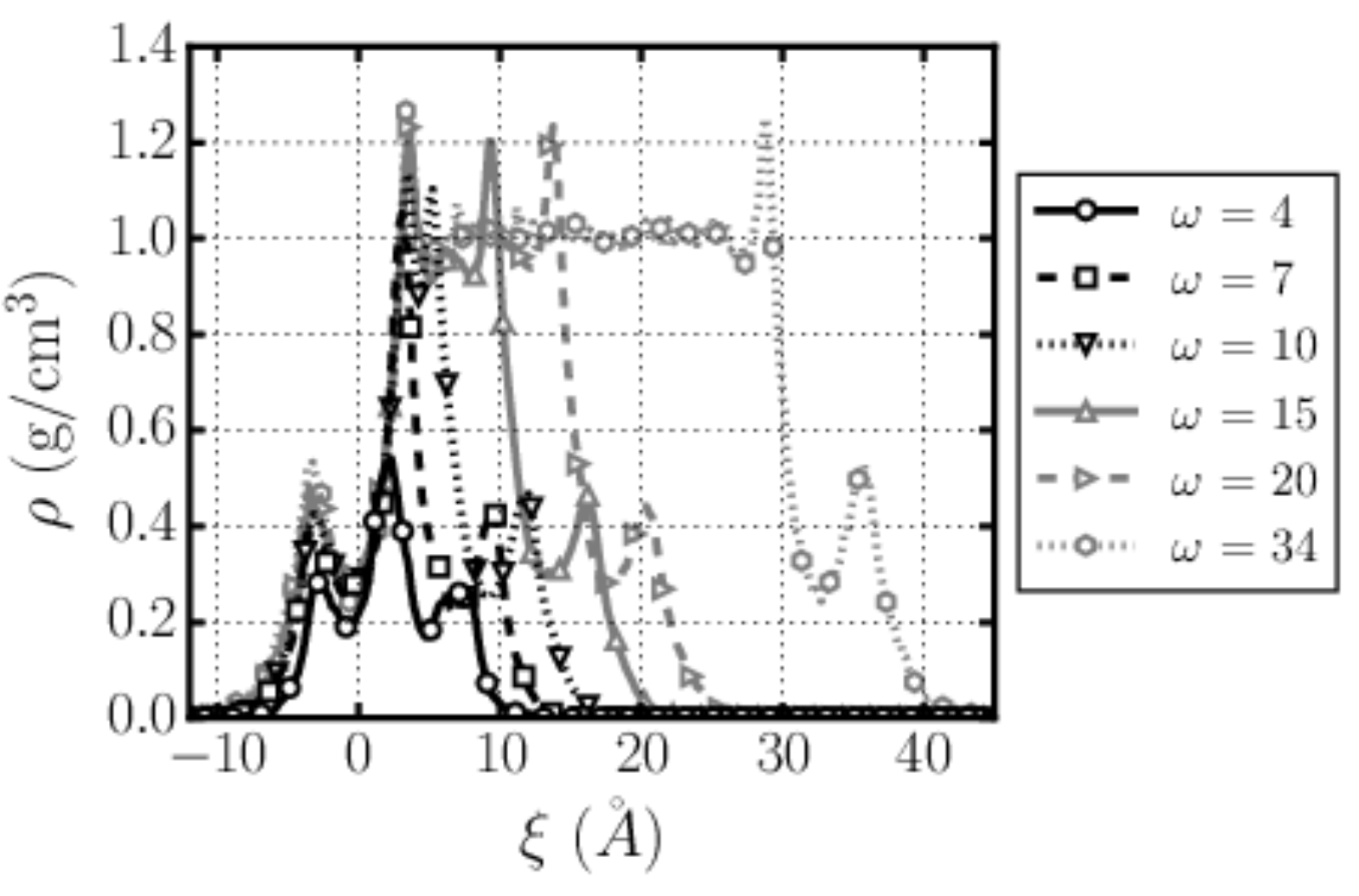}
\caption{Water density profile for stacked DMPC phospholipid bilayers at different hydration levels:  $\omega = 4$ (dots),  $\omega = 7$ (squares), $ \omega = 10$ (down triangles), $\omega = 15$ (up triangles), $\omega = 20$ (side triangles), and $\omega = 34$ (hexagons). Symbols are placed to identify the corresponding lines. Adapted from (Calero et al. 2016).}
\label{Fig:watdensprofile}
\end{center}
\end{figure}

\subsubsection{Water density profile for different hydration levels}

By adopting the  local distance  $\xi$ to the membrane, it is possible to study 
the structure of the water-membrane interface, represented by the density profile as a function of  $\xi$  (Figure \ref{Fig:watdensprofile}).  The water density profile for stacked membranes at different hydration level emphasizes structural changes in the way water molecules locate themselves at the interface. From inspection of the water density profile for the fully hydrated membrane (with hydration level = 34) we can clearly distinguish three main regions: an inner (interior) layer ($\xi < 0$), the first hydration layer ($0 < \xi < 5\AA$), and outer layers ($\xi > 5 \AA$) (Pandit et al. 2003). 

For the other less hydrated cases the same structure is preserved, although the exterior layer is thinner (in the cases with hydration $\omega =$ 15, 20) or nonexistent (in the cases with hydration level $\omega =$ 4, 7, 10). In addition, the density profiles in Figure \ref{Fig:watdensprofile} show that, as hydration increases, water molecules accumulate in a layering structure. Indeed, water molecules first fill the interior and first hydration layer before starting to accumulate in the outer region ($\xi > 5 \AA$). The interior and the first hydration layer become saturated when $7 < \omega < 10$, in agreement with X-ray scattering experiments (Kuv\u{c}erka et al. 2005). 

For the least hydrated cases (hydration levels $\omega  =$ 4, 7), Calero et al. (Calero et al. 2016) observe that, although not yet ``full'', the inner region and the first hydration layer are occupied. This implies that the inner water is a {\it structural} part of the membrane and remains even if the hydration is low. As will be discussed in the following, inner water can be seen as an essential component of the membrane that plays a structural role with hydrogen bonds that are forming bridges between lipids
(Pasenkiewicz-Gierula et al. 1997, Lopez et al. 2004).

\subsubsection{Structure: Hydrogen bonds}

The structure and dynamics of the hydrogen bond network regulates the dynamical and thermodynamical properties of liquid water  (Bianco and Franzese 2014, de los Santos and Franzese 2009, 2011, 2012, Franzese and de los Santos 2009, Franzese et al. 2010, Franzese and Stanley 2002, Kumar et al. 2006, 2008, Laage and Hynes 2006, Mazza et al. 2011, 2012, 2009, Stanley et al. 2009, Stokely et al. 2010), as well as 
the properties of hydrated biological macromolecules (Bianco and Franzese 2015, Bianco, Franzese, Dellago and Coluzza 2017, Bianco et al. 2012, Bianco, Pag\`es-Gelabert, Coluzza and Franzese 2017, Franzese et al. 2011, Vilanova et al. 2017).
Here, we analyze in detail the case of the hydrogen bond network formed by water molecules between stacked membranes and we discuss its relation to the dynamics of the hydration water. 

From the trajectories of all-atom MD simulations Calero et al. (Calero et al. 2016) evaluate the number of hydrogen bonds formed by water  between membranes and its dependence with the hydration level. They adopt the widely employed geometric definition of the hydrogen bond: two molecules are hydrogen bonded if the distance between donor and acceptor oxygen atoms satisfies $d_{OO} < 3.5$~\AA~ and the angle formed by the OH bond of the donor molecule with the OO direction is $\theta < 30^o$. In addition to hydrogen bonds formed between water molecules they also consider hydrogen bonds created between water and phosphate or ester groups of the DMPC phospholipid (Bhide and Berkowitz 2005).  

\begin{figure}%[H]
\begin{center}
\vspace*{0.5cm}
\includegraphics*[angle=0, width=11cm]{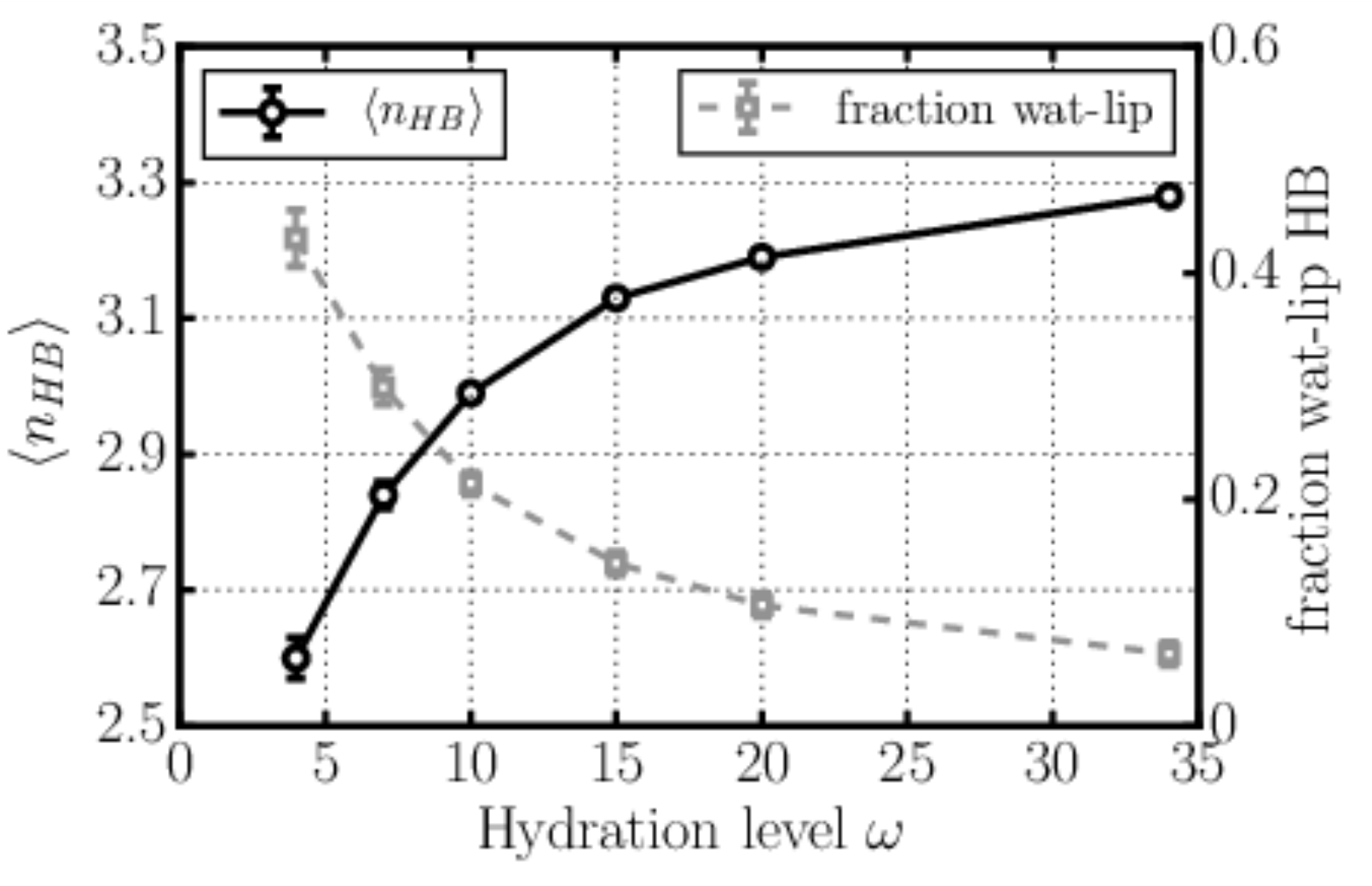}
\caption{Hydrogen bonds formed
  by water molecules confined in stacked phospholipid membranes as a
  function of their hydration level.
   The average total
  number of hydrogen bonds (black circles) increases, while 
the average fraction of hydrogen bonds that water 
  forms with lipid groups (gray squares) decreases, when $\omega$ increases.
  Adapted from  (Calero et al. 2016). } 
\label{Fig:avHBhyd}
\end{center}
\end{figure}

The average number $\langle n_{HB}\rangle$ of hydrogen bonds formed by water molecules depends on the hydration level of the membrane (Figure~\ref{Fig:avHBhyd}). The number of water-water hydrogen bonds decreases monotonically as the membrane hydration $\omega$ is reduced. In contrast, the fraction of hydrogen bonds formed by water with selected groups of the lipid increases for decreasing $\omega$, amounting to almost half of all the hydrogen bonds at low hydration  ($\omega = 4$).

\begin{figure}[h!]
\begin{center}
\vspace*{0.cm}
\includegraphics*[angle=0, width=11cm]{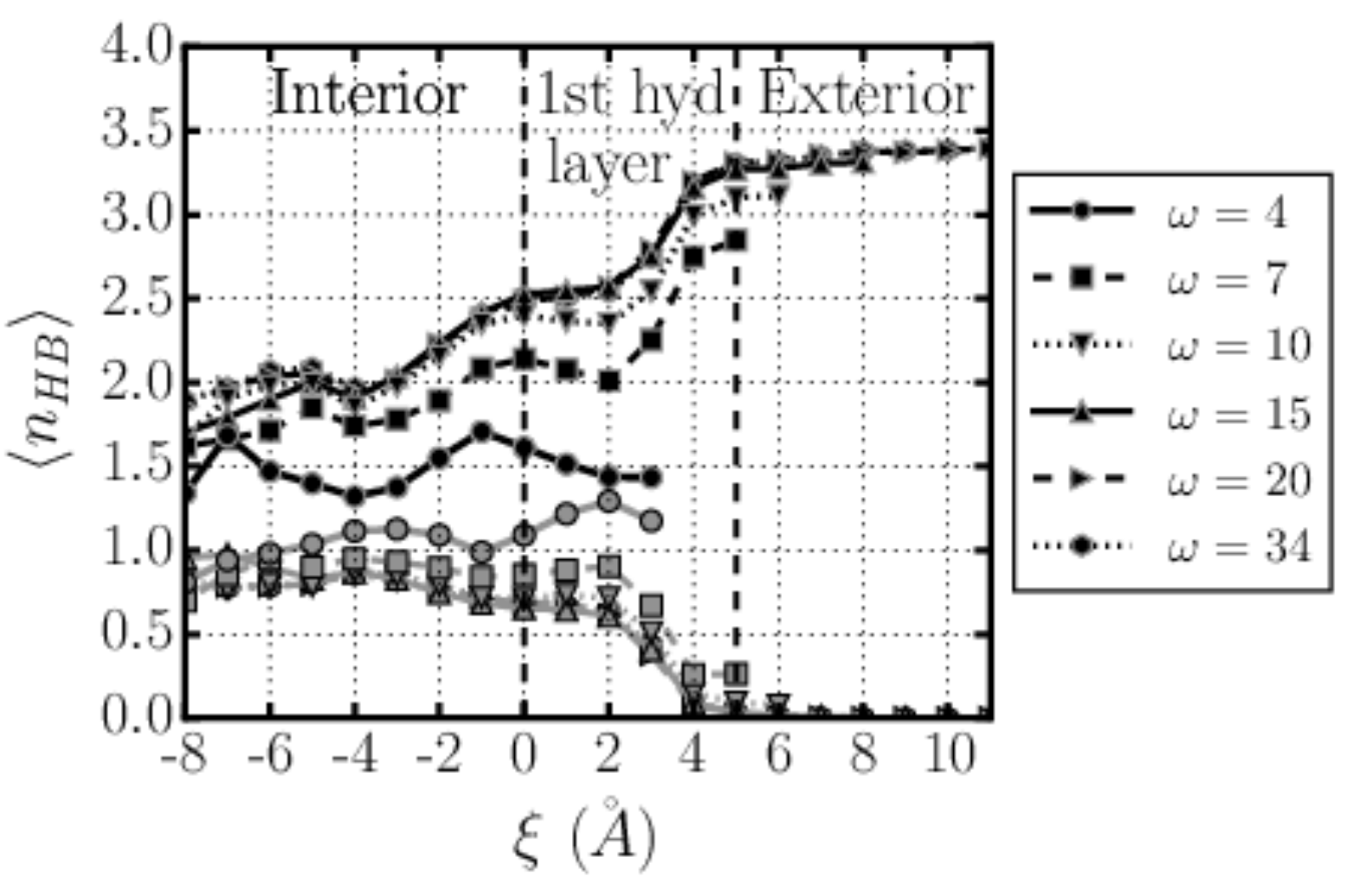}
\caption{Average number of hydrogen bonds $\langle n_{HB} \rangle$ as a function of $\xi$ for  different hydrations: between water molecules (black) and between water molecules and selected groups of the phospholipid (gray). Adapted from (Calero et al. 2016).}
\label{Fig:Hbondprofile_layers}
\end{center}
\end{figure}

Due to the importance of hydrogen bonds in water properties, the hydrogen bond profile provides a good measure of the structure of the water-membrane interface. The average number of hydrogen bonds $\langle n_{HB} \rangle$ formed by water molecules shows a complex  dependence on distance $\xi$ of water to the membrane 
 with different hydration level (Figure~\ref{Fig:Hbondprofile_layers}) .
For the completely hydrated membrane (with $\omega = 34$), the number of water-water hydrogen bonds is $\approx 3.45$ in the exterior part of the membrane, decreases in the first hydration layer, and becomes $\approx 2$ in the interior part of the membrane. The number of water-lipid hydrogen bonds is, by necessity,  zero at the exterior region of the interface, increases in the first hydration layer and  becomes $\approx 1$ in the interior part of the membrane. The same profile is observed approximately for all hydration levels, except for the two least hydrated cases with $\omega=4, 7$. In those cases, the number of water-water hydrogen bonds is lower, whereas the number of water-lipid hydrogen bonds is higher than in cases where the interior and first hydration layers of the membrane are saturated with water molecules.

\begin{figure}[h!]
\begin{center}
\vspace*{0.5cm}
\includegraphics*[angle=0, width=8.5cm]{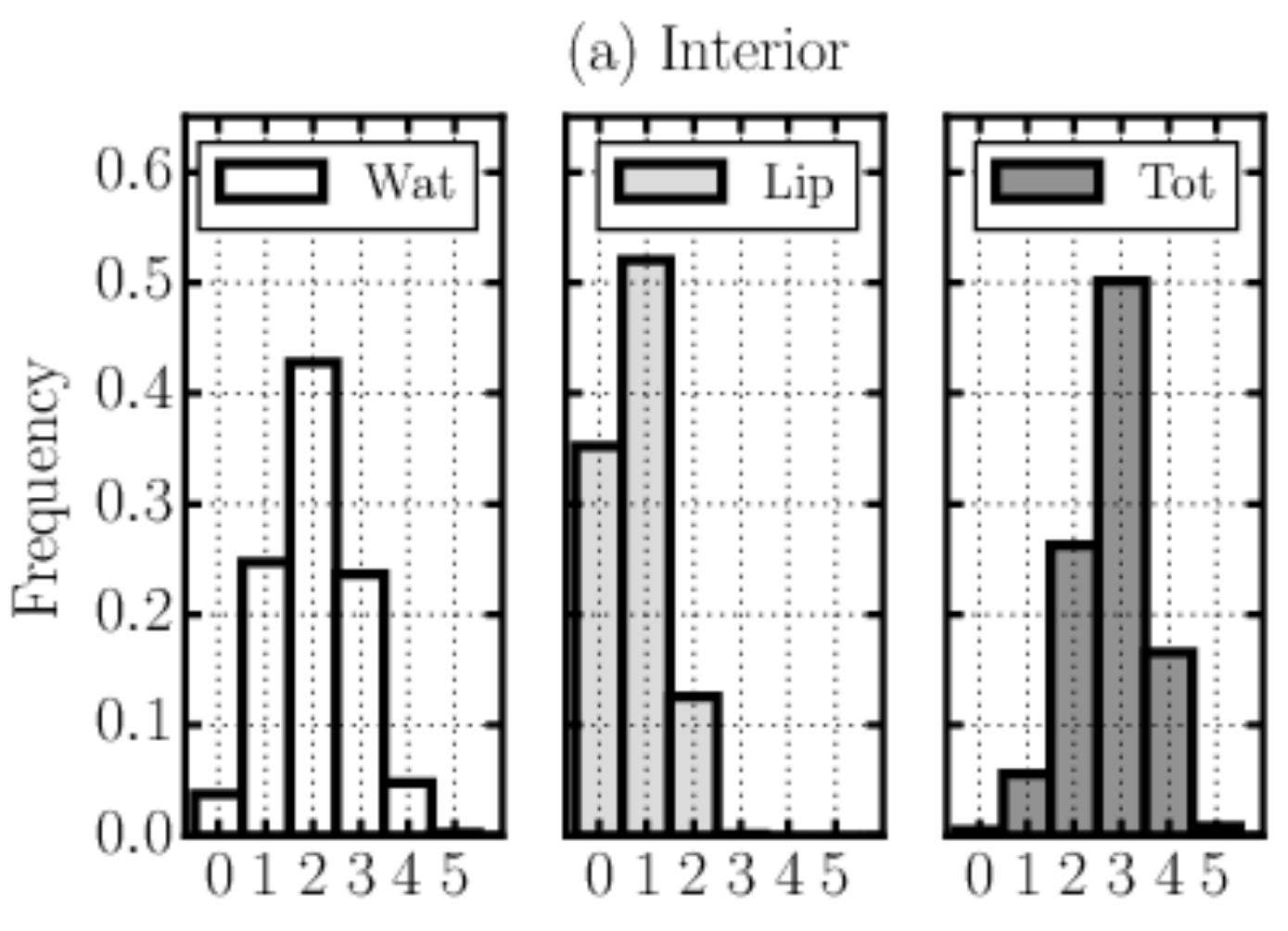}
\includegraphics*[angle=0, width=8.5cm]{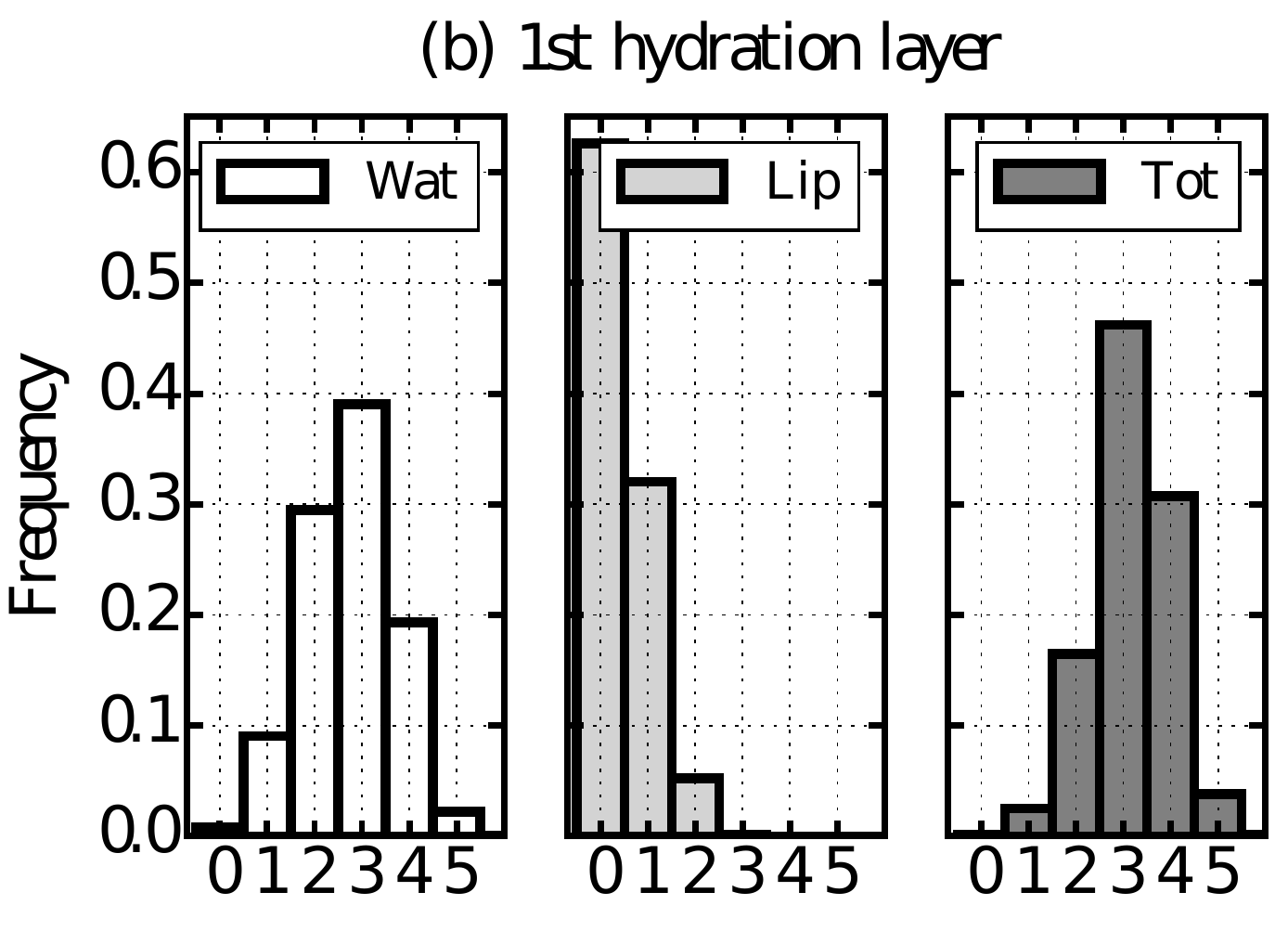}
\includegraphics*[angle=0, width=8.5cm]{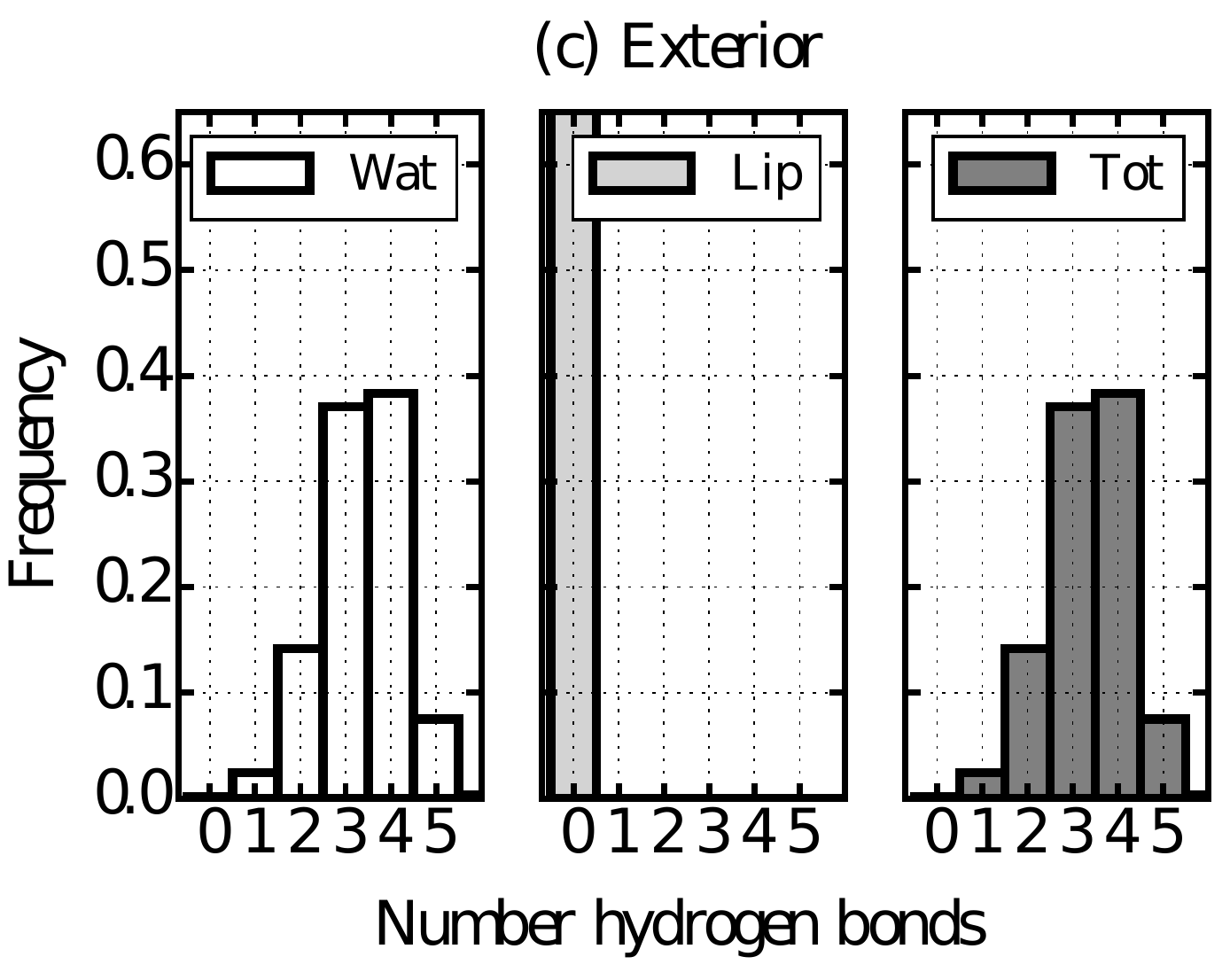}
\caption{Distribution of the number of hydrogen bonds formed by a water molecule in a completely hydrated ($\omega=34$) stacked membrane at different regions of the interface: (a) Interior ($\xi < 0$), (b) 1st hydration layer ($0 <\xi < 5\AA$), and (c) Exterior ($\xi > 5\AA$), showing the distribution of  water-water  (Wat), water-lipid  (Lip) and total (Tot) hydrogen bonds. The probability distribution of hydrogen bonds is different in each region.}
\label{Fig:distrHB_layers}
\end{center}
\end{figure}

To further understand the structure of the water-membrane interface, we calculate here
the probability distribution of the number of hydrogen bonds formed by water molecules located at different regions of the interface for the completely hydrated membrane ($\omega=34$). We find that  the probability distribution of hydrogen bonds is different in each region (Figure~\ref{Fig:distrHB_layers}). 

Water has preferentially two water-water hydrogen bonds in the inner part of the membrane, three in the first hydration shell, and four in the outer part.
At the same time, a water molecule forms often one, and seldom two hydrogen bonds with the lipid in the inner part, while in the first hydration shell it forms one with significant probability and very rarely two, although most water does not bind to the phospholipid. However, as discussed by  Calero et al. (Calero et al. 2016), even at low hydration there is no unbound water near the membrane. 

This result is 
at variance with the hypothesis of {\it free}, fast water in weakly hydrated phospholipid bilayers that has been proposed  to account for rapidly relaxing signals associated with the reorientation of water molecules in experiment  (Tielrooij et al. 2009, Volkov et al. 2007).  As discussed elsewhere (Calero and Franzese 2018), the experiments can be interpreted better by analyzing how the water dynamics changes as a function of distance to membrane, showing that it is possible to identify the existence of an interface between the first hydration shell,  extremely slow and partially made of water bonded to the membrane, and the next shells, faster but still one order of magnitude slower than bulk water.

\section{Effect of temperature on the hydration water dynamics}

Next we study how hydration water dynamics changes with temperature $T$.
We simulate a fully hydrated ($\omega = 34$) DMPC phospholipid membrane using the NAMD simulation package and study how the relaxation of the rotational dipolar correlation function $C_{\rm sim}^{\rm rot}(t)$, defined in Eq.(\ref{Eq:Crot}), varies with $T$. 

In particular, we carry out 10 independent simulations, with production runs of 20 $ps$ each, in the $NVT$ ensemble at different temperatures and at an average pressure $P$ of 1 atm, following the same simulation protocol described in (Calero et al. 2016).
. 

To simplify our analysis, we analyze the dynamics of hydration water by grouping the molecules based on their average distance from the membrane at the beginning of each run,  regardless their following trajectories. This distance definition is different from the distance $\xi$ presented in section \ref{subIIIA1}, but, as we will show next, is able to quantify the effect of temperature changes within our 20 $ps$ simulation and in the range of temperatures we consider here.
 
\begin{figure}[h!]
\begin{center}
\vspace*{0.5cm}
\includegraphics*[angle=0, width=16cm]{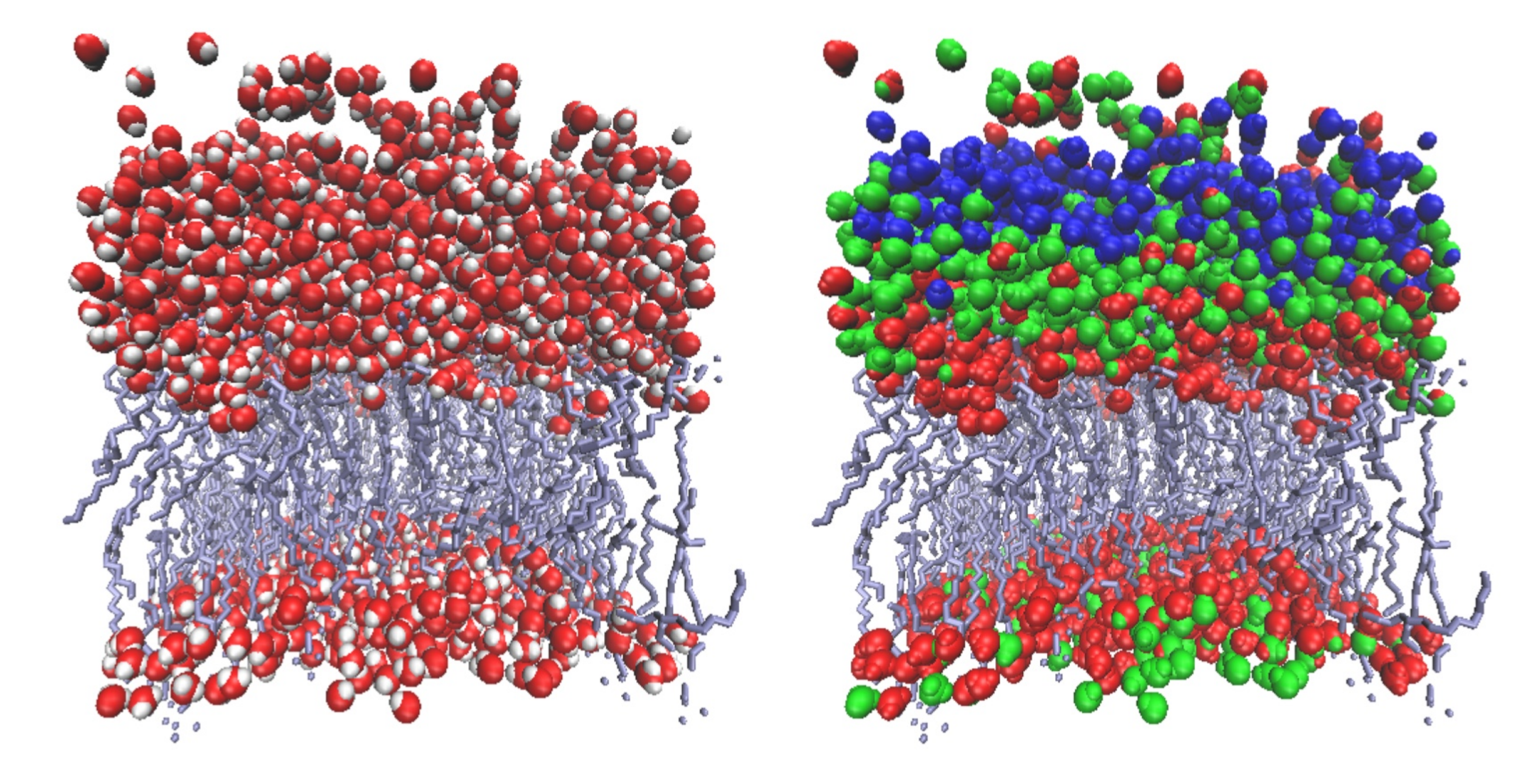}
\caption{Snapshot of the hydrated DMPC phospholipid membrane, represented by the grey carbon backbone of the lipids. The system has periodic boundary conditions, reproducing water between stacked membranes. 
(Left panel)  An initial configuration with water oxygens represented in red and water hydrogens in white. (Right panel) The same configuration where now oxygen and hydrogen atoms are colored based on their distance form the closest lipid atom: (1) red for those at distance 
up to 3 \AA, (2) green for those at larger distances up to 9 \AA,  (3) 
blue for the rest in (9-15] \AA range.  Careful inspection reveals that there are water molecules with atoms with different colors. These molecules across regions are excluded from our analysis.}
\label{Fig:3-layers}
\end{center}
\end{figure}

At the beginning of each run we classify water molecules into 
 three different groups: (1) those within a distance of up to 3 \AA~from any phospholipid atom, including those between the lipids; (2) those at larger distances up to  9 \AA~from the lipids, and (3) those at distances larger than 9 and up to 15 \AA~from the lipids (Figure~\ref{Fig:3-layers}).  Because the classification is atom-based, there is a finite probability of having molecules  that initially are between two regions. For sake of simplicity, these molecules are excluded from the analysis.

\begin{figure}[h!]
\begin{center}
\vspace*{0.5cm}
\includegraphics*[angle=0, width=12cm]{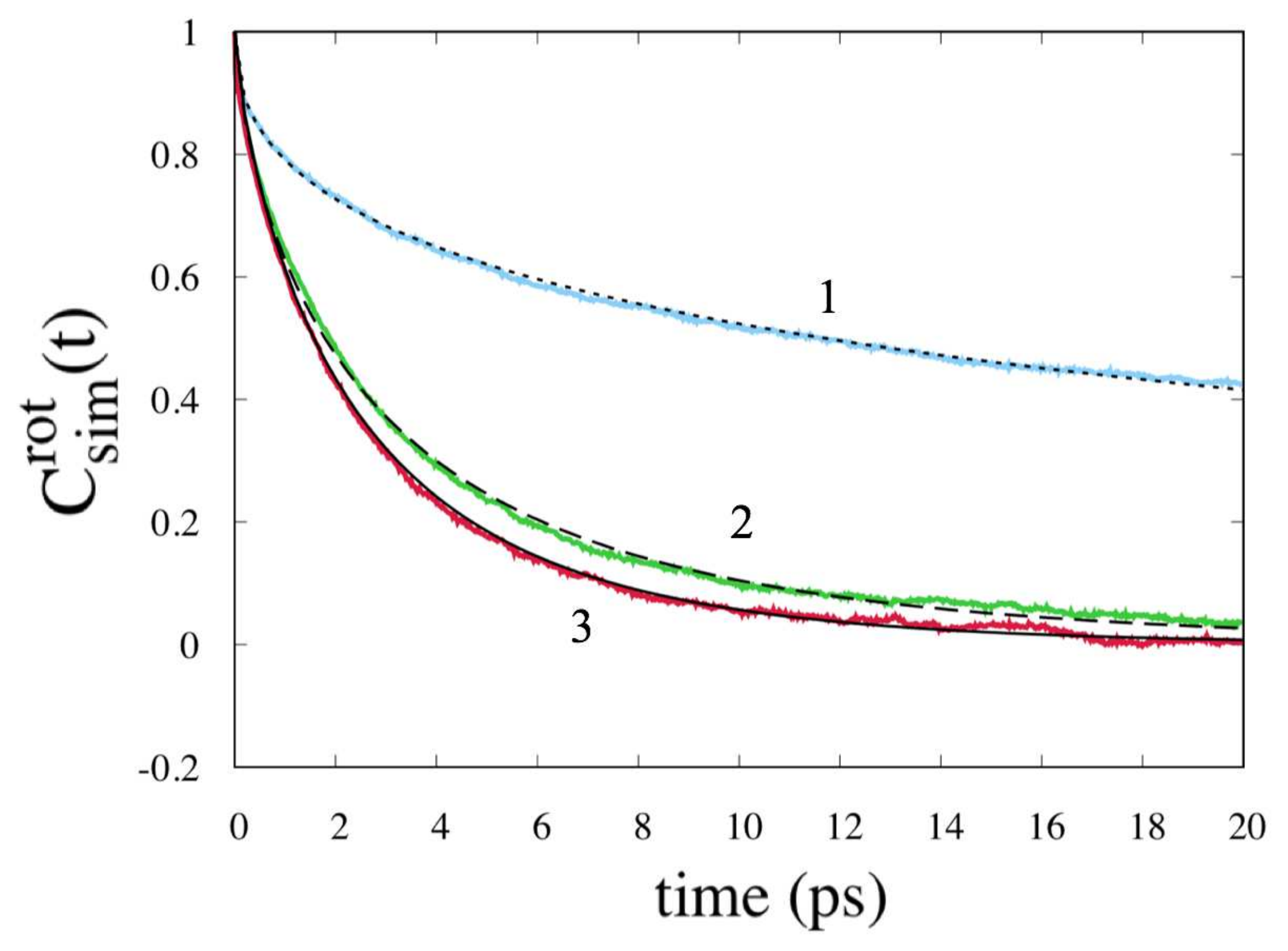}
\caption{Rotational dipolar correlation function  $C_{\rm sim}^{\rm rot}(t)$  as function of time from MD simulations of water between stacked membranes at $T$=288.6 K and 1 atm. From top to bottom: Correlation of (1) water molecules within 3 \AA~ average distance from the membrane, (2) at  larger distances up to 9 \AA, (3) at distances above 9 and up to 15 \AA. Data from simulations are represented as colored points (blue for 1, green for 2, red for 3), fit with 
stretched exponential decay Eq.(\ref{C_fit}) with continuous lines (dotted for 1, dashed for 2, solid for 3). Fitting parameters are given in Table \ref{table:paramT288.6}. }
\label{Fig:low-T}
\end{center}
\end{figure}

We simulate the systems for temperatures between 303 K and 288.6 K, at which  the membrane remains in the liquid phase, finding a clear slowing-down of the  rotational dipolar correlation function $C_{\rm sim}^{\rm rot}(t)$ as the temperature decreases. To quantify this effect, at each $T$  we fit $C_{\rm sim}^{\rm rot}(t)$
with  the Kohlrausch-Williams-Watts stretched exponential function, typical of glassy systems  (Fierro et al. 1999, Franzese 2000, 2003, Franzese and Coniglio 1999, Franzese
et al. 1999, Franzese and de los Santos 2009, Franzese et al. 1998, Kumar et al. 2006),
\begin{equation}
 C_{\rm sim}^{\rm rot}(t) = \exp \Bigg[ - \bigg( \frac{t}{\tau} \bigg) ^{\beta} \Bigg],
\label{C_fit} 
\end{equation} 
where $\tau$ is the characteristic relaxation time and $\beta$ is the stretched exponent 
(Figure~\ref{Fig:low-T}, Table \ref{table:paramT288.6}). Note that $\beta$ is always larger than 1/3, as expected from general considerations (Fierro et al. 1999, Franzese and de los Santos 2009, Franzese et al. 1998) and that $\tau$ is comparable to the rotational correlation function of bulk water at 250 K (Kumar et al. 2006), that is $\approx $38 K less than the temperature of the hydrated membrane considered here. 

\begin{table}%[H]
\small 
\centering
\begin{tabular}{|c|c|c|}
\hline
Parameter  & $\beta$  & $\tau$ $(ps)$   \\
\hline
\hline
Group 1 & 0.440 $\pm$ 0.002 & 26.9 $\pm$ 0.1 \\ 
\hline
Group 2 & 0.685 $\pm$ 0.003 & 3.05 $\pm$ 0.06 \\ 
\hline
Group 3 & 0.767 $\pm$ 0.002 & 2.53 $\pm$ 0.02 \\ 
\hline
\end{tabular}
\caption{Stretched exponential decay fitting parameters from Eq.(\ref{C_fit}) for the MD data of rotational dipolar correlation function of water between stacked membranes at $T$=288.6 K and 1 atm for molecules in  groups at different average distances from the membrane:  1, 2, and 3, as defined in the text.}
\label{table:paramT288.6}  
\end{table}

\begin{table}%[H]
\small 
\centering
\begin{tabular}{|c|c|c|c|c|}
\hline
 $T$ (K) & 303.0 & 295.8 & 292.2 & 288.6  \\
\hline
\hline
$\tau_{1}$(ps) &  15 $\pm$ 1  &  18 $\pm$ 1 &  20 $\pm$ 1  & 26.9 $\pm$ 0.1  \\ 
\hline
$\tau_{2}$(ps) & 2.0 $\pm$ 0.5 & 2.5 $\pm$ 0.2 & 2.8 $\pm$ 0.2 & 3.05 $\pm$ 0.06  \\ 
\hline
$\tau_{3}$(ps) & 1.5 $\pm$ 0.2 &  2.0 $\pm$ 0.1 & 2.2 $\pm$ 0.1 & 2.53 $\pm$ 0.02  \\ 
\hline
\end{tabular}
\caption{Rotational relaxation time estimated as the time at which  $C_{\rm sim}^{\rm rot}(t)$ reaches the value $1/e$ for different temperatures $T$ 1 atm, for molecules in  groups 1, 2, and 3, as defined in the text. For the lowest temperature  $T=288.6$ K we use the times from Table \ref{table:paramT288.6}, that have approximately the same values as the estimate for $1/e$, but a smaller error.}
\label{table:tau-vs-T}  
\end{table}

 We observe a much slower decay for the water molecules of the first group (up to 3 \AA) compared to that of the molecules found in the second (3 - 9 \AA) and third (9 - 15 \AA) groups. The characteristic relaxation times obtained for each case increase for decreasing $T$  (Table \ref{table:tau-vs-T}). We find that the correlation time $\tau_1$ estimated for $1/e$ at $T=303$ K for the group  1 of water molecules at the closest distance from the membrane is approximately equal to the integral correlation time  Eq.(\ref{tau}) at the same temperature and full hydration,  ($12.4 \pm 0.3$) ps, and that  $\tau_1$ increases for decreasing $T$.
 
To make a better comparison of the relaxation times estimated here with those calculated in section \ref{sec:tau}, we calculate the integral rotational correlation time from Eq.(\ref{tau}) for the fitted functions of $C_{\rm sim}^{\rm rot}(t)$ for $T=288.6$ K. We find: $\tau_{\rm rot,1} = 70.1 $ ps, $\tau_{\rm rot,2} = 3.9 $ ps, and $\tau_{\rm rot,3} = 3.0 $ ps for the groups 1, 2 and  3 of water molecules, respectively. As expected the integral correlation time for the group  of water hydrating the membrane within 3 \AA~ is much larger than the integral correlation time  for $T=303$ K at full hydration, ($12.4 \pm 0.3$) ps, consistent with the slowing down of the system for decreasing $T$.

We observe that our preliminary analysis of the rotational correlation times shows  $T$-dependence also  for water in groups 2 and 3, at distance larger that 3 \AA~ from the membrane, but with only a minor difference between the two groups. This results seems to suggest that the effect on water of the soft interface  is only minor at a distance larger than 3 \AA~ from the membrane. In the next section we will discuss if this is really the case.

\section{Extension of the structural effect of the membrane on interfacial water}

In order to investigate the effect of the membrane-water interactions to the structural properties of water, as in  (Martelli, Ko, Borallo and Franzese 2018), we
employ a recently-introduced sensitive local order metric (LOM)  that  maximizes the spatial overlap between a local snapshot and a given reference structure (Martelli, Ko, O\u{g}uz and Car 2018). 

The LOM measures and grades the degree of order present in the neighborhood of an atomic or molecular site in a condensed medium. It is endowed with high resolving power and high flexibility 
(Martelli, Giovambattista, Torquato and Car 2018, Martelli, Ko, O\u{g}uz and Car 2018), and allows one to look for specific ordered domains defined by the location of selected atoms (e.g., water oxygens) in a given reference structure, for which it is found the best alignment out of a sufficiently large number of  orientations picked at random and with uniform probability.
 Typically, the reference  is taken to be the local structure of a perfect crystalline phase. From the LOM averaged over all sites in a snapshot, one obtains a score function $S$, representing an order parameter that tends to $0$ for a completely disordered system, and is $1$ for a system perfectly matching the reference structure. 

\begin{figure}[h!]
\begin{center}
\vspace*{0.5cm}
\includegraphics*[angle=0, width=8cm]{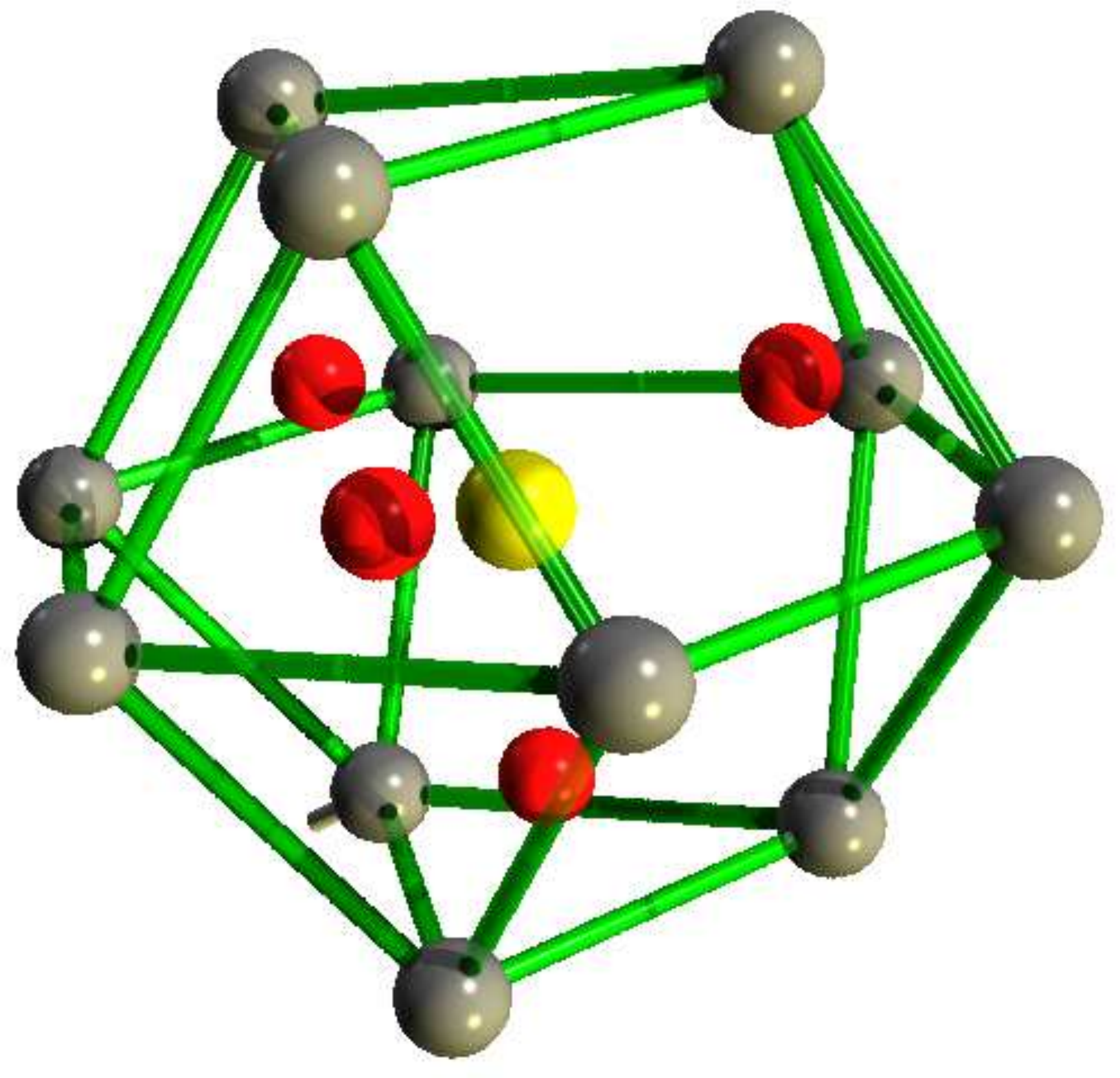}\\
\caption{Pictorial representation of the second neighbor shell
in cubic ice (or cuboctahedron). Gray spheres indicate the position of oxygen atoms, while the green lines emphasize the geometrical structure. The central oxygen is depicted by the yellow sphere, while the red spheres represent the first shell of neighboring oxygens.}
\label{Fig:cuboct}
\end{center}
\end{figure}

As in Martelli et al.  (Martelli, Ko, Borallo and Franzese 2018), we inspect the intermediate-range order (IRO) of water employing, as a reference structure, the configuration of the oxygens of cubic ice (Ic) at the level of the second shell of neighbors. Similar results hold when adopting the second shell of neighbors of hexagonal ice. The structure described by the $12$ second neighbors in Ic (Figure~\ref{Fig:cuboct}), define the Archimedean solid cuboctahedron (Cromwell 1999). 

\begin{figure}[h!]
\begin{center}
\vspace*{0.5cm}
\includegraphics*[angle=0, width=13cm]{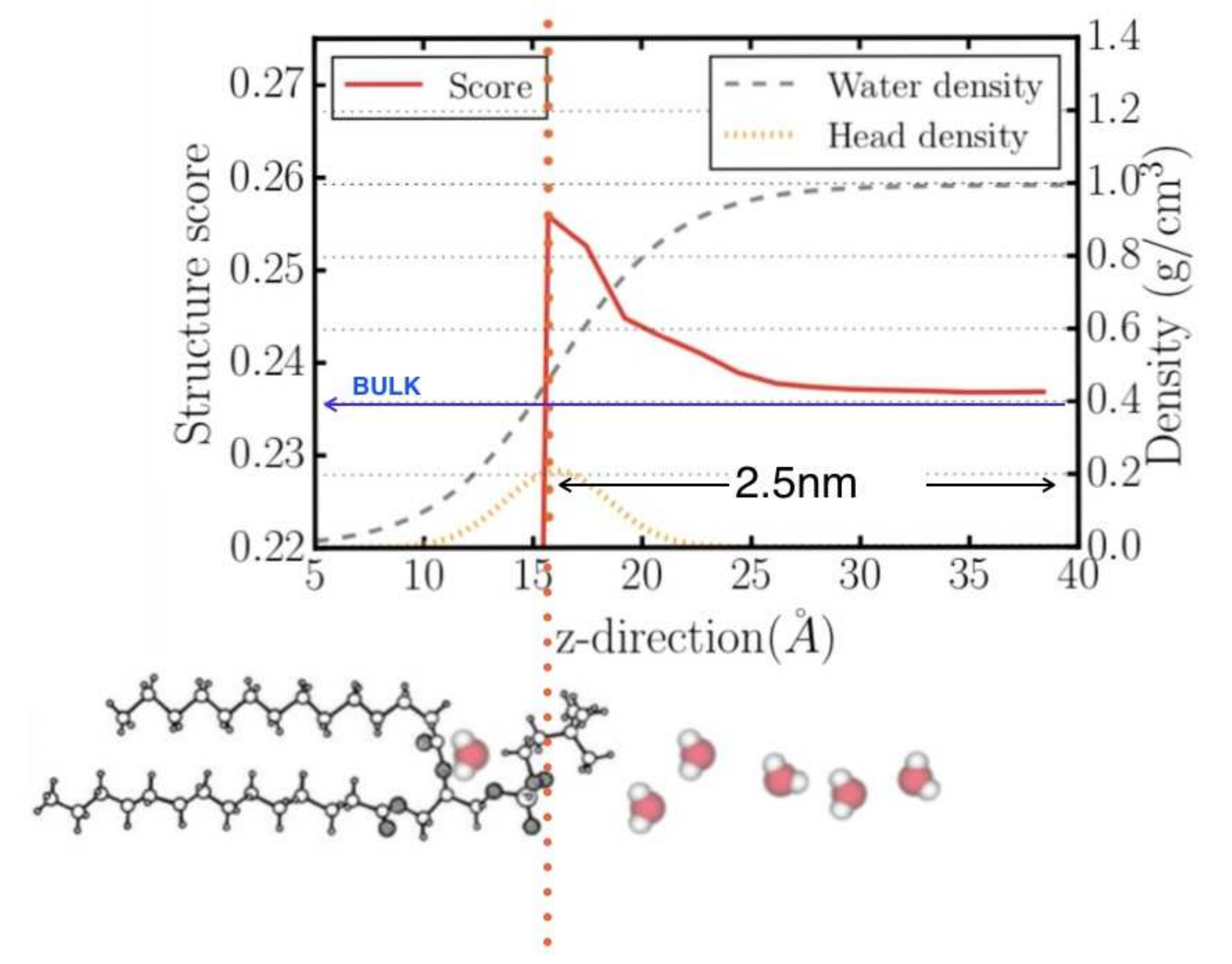}
\caption{Upper panel: Structure of water and lipids along the $z$-axis perpendicular to the average surface plane of the membrane. The right vertical axes of the plot represents the 
density profile of lipid heads (in orange) and water (in grey) along $z$. The $z$ of maximum heads density coincides approximately with the largest variation of water density and is marked by a vertical dotted line in red. The left vertical axes represents the average score $S$ of water (in red) along $z$. $S$ is zero inside the membrane and is higher than its bulk value (blue arrow) within at least 2.5 nm distance from the surface. 
Lower panel: Schematic representation of  a configuration with a lipid head near the surface location (vertical dotted line in red) and water molecules mimicking their density profile as indicated in the upper panel..}
\label{Fig:score}
\end{center}
\end{figure}

With this choice, we calculate
 the profile of the score $S$  as a function of the  distance from the membrane surface on the $z$-axis perpendicular to the surface. The membrane surface here is defined as the place where the density of  lipid heads reaches a maximum along the $z$-axis perpendicular to the average surface plane and approximately coincides with the distance where the water density has the largest variation along $z$ and an approximate value of half the bulk density (Figure~\ref{Fig:score}). 
 
The first observation is that the interface is smeared out over $\approx$1 nm as a consequence of the average over the membrane fluctuations, with the water density changing smoothly from 0 to the bulk value, 1 g/cm$^3$. The water density profile suggests no membrane-induced structure in the hydration water. 

Nevertheless, we find that  $S$ approaches, but never reaches the value of $S$ in bulk liquid water at the same thermodynamic conditions, indicating that, at a distance as far as $\simeq 2.5$ nm from the surface, the structural properties of bulk water are not recovered yet (Figure~\ref{Fig:score}).

\begin{figure}[h!]
\begin{center}
\vspace*{0.5cm}
\hspace{-10cm}a)\\
\includegraphics*[angle=0, width=14cm]{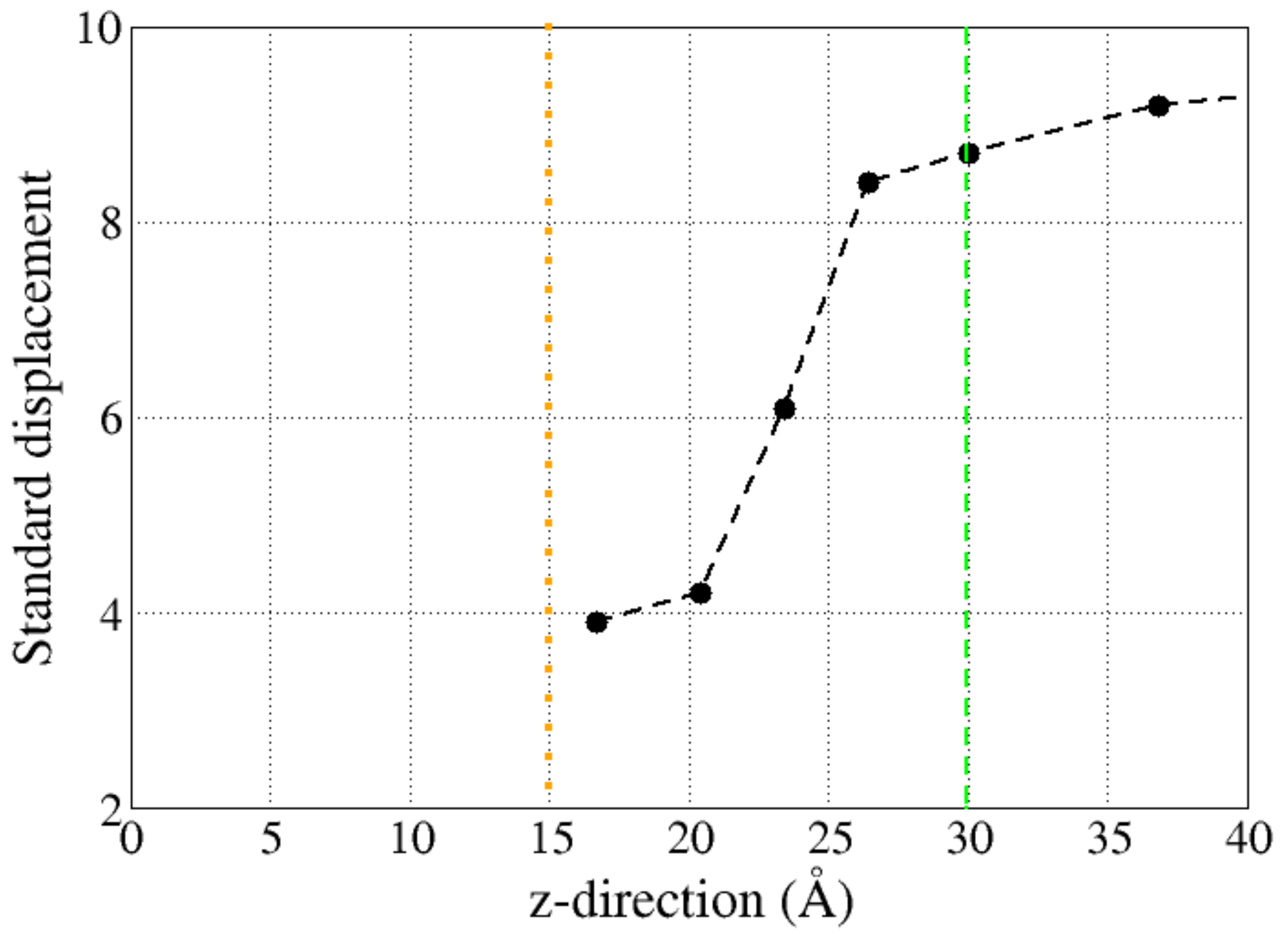}\\
\caption{Standard displacement of water molecules near the membrane as a function of the distance $z$ from the membrane.  The standard displacement largely decrease  where the  variations in $S$ with respect to bulk  is appreciable (region between the two dotted vertical lines).
The distance is calculated with respect to the center of the bilayer ($z=0$).
Adapted from (Martelli, Ko, Borallo and Franzese 2018).}
\label{Fig:BU}
\end{center}
\end{figure}

On the other hand, Martelli et al.  (Martelli, Ko, Borallo and Franzese 2018) show that dynamical bulk properties are recovered at much shorter distances. 
To characterize the translational dynamics of water at the interface over a $1$ ns-long simulation, they calculate the average O--O distances  from the first minimum of the O--O radial distribution function of bulk water, and for each water near the membrane they evaluate the ``standard displacement'', i.e. the distance traveled in units of  O--O distances (Figure~\ref{Fig:BU}).

At the considered thermodynamic conditions, bulk water would travel a standard displacement between 8 or 10 in a  run of $1$ ns. This is consistent with the observed quantity for water between 1 nm and 2.5 nm from the membrane (Figure~\ref{Fig:BU}).

By approaching the membrane surface, instead, water slows down considerably and the standard displacement reduces to $\simeq 4$,
similar to the displacement in deeply undercooled liquid water 
(Martelli, Ko, O\u{g}uz and Car 2018). This value is quite small, considering that a standard displacement 
$\lesssim 1$
would correspond to a system in which translational degrees of freedom are frozen and water molecules would only rattle in their nearest neighbors cages.
Such decrease occurs synergistically with a considerable increase of $S$ (Fig.~\ref{Fig:score}), suggesting that the structuring in the IRO can be ascribed to the significant slowdown of translational degrees of freedom. 

\begin{figure}[h!]
\begin{center}
\vspace*{0.5cm}
\hspace{-10cm}a)\\
\includegraphics*[angle=0, width=14cm]{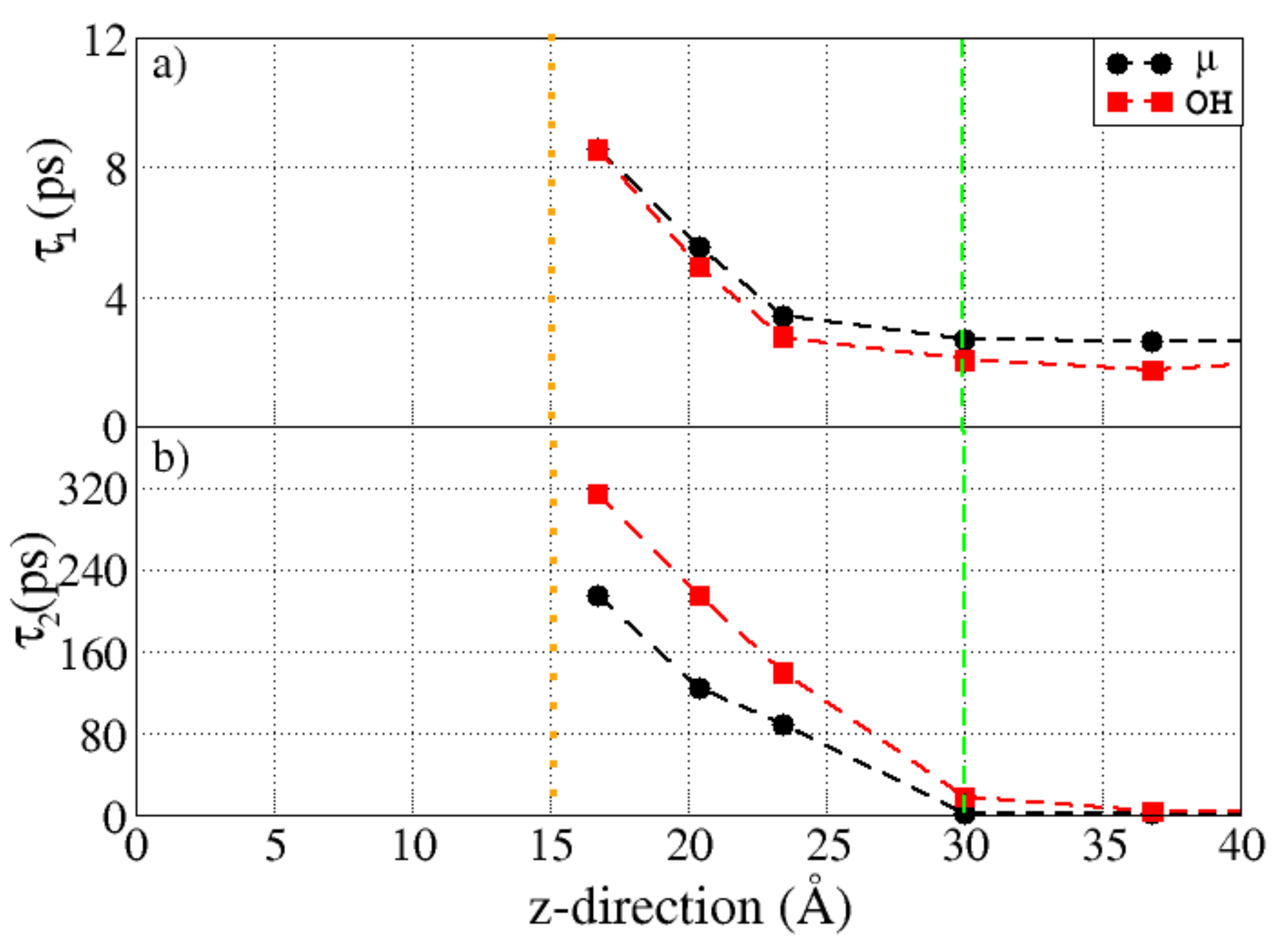}\\
\caption{Relaxation times 
(a)  $\tau_1$ 
and 
(b) $\tau_2$
for the dipole (black circles) and OH (red squares) vector 
correlation functions as a function of the distance from the membrane surface. 
All the relaxation times, with $\tau_2\gg \tau_1$,  rapidly increase  
 where the  variations in $S$ with respect to bulk  is appreciable (region between the two dotted vertical lines).
The distance is calculated with respect to the center of the bilayer ($z=0$).
 Adapted from (Martelli, Ko, Borallo and Franzese 2018).}
\label{Fig:tau}
\end{center}
\end{figure}

Based upon the observation that DMPC heads contains both P- and N- headgroups, with different charges, it is conceivable to imagine that P-OH and N-O interactions among lipids and water have different strengths and, ultimately, that the P- and the N- headgroups contribute differently to the structural properties of water close to the membrane surface. In order to quantify this effect, Martelli et al. (Martelli, Ko, Borallo and Franzese 2018)
 compute the dipole and the $\vv{OH}$ correlation functions. They fit the correlation functions with a double exponential function, obtaining two characteristic relaxation times, $\tau_1$ and $\tau_2$, with  $\tau_2\gg \tau_1$ (Figure~\ref{Fig:tau}). 
 
 The rotation of water considerably slows down as water molecules approach the membrane. In
particular, the distances at which the change is more appreciable are the same as those for the translations and for the increase in the score function $S$.
On approaching the lipid membrane, both $\tau_1$ and $\tau_2$ increase and, interestingly, the relaxation times for
$\vv{OH}$ increase at a pace higher than the relaxation times for $\vv{\mu}$, indicating that the P--HO interaction is stronger than the N--O interaction. 

The slowing down of the rotational dynamics decreases as water moves away from the interface.
At distances $\geq 1.5$ nm, the relaxation times are almost indistinguishable and, as one would expect in bulk liquid water, $\tau_{1}^{\mu}\simeq\tau_{2}^{\mu}\simeq \tau_{1}^{OH}\simeq\tau_{2}^{OH}$ (Figure~\ref{Fig:tau}).

\section{Conclusions}\label{Sect__Conclusions}

The dynamics and the structure of a membrane and the hydration water are related in a way that  goes beyond the  hydrophobic effect that allows the membrane to self-assemble.
MD simulations of water confined between stacked phospholipid bilayers at different hydration levels reveal that, adopting  
an appropriate local distance of water from the membrane (Berkowitz et al. 2006, Pandit et al. 2003),  water forms layers around the membrane (Calero et al. 2016). These layers are not observed on the macroscopic scale because 
the spatial fluctuations  of the membrane smeared them out.
 Nevertheless, the local distance, and the analysis at different hydration levels, allow to identify different dynamics behavior of the layers clearly.

In particular, the closer is water to the membrane, the slower is its translational and rotational dynamics, being always slower than bulk water (Calero et al. 2016, Martelli, Ko, Borallo and Franzese 2018). These results are putting into  question
the possible existence of rotationally {\it fast} water molecules near the membrane, as  proposed   to rationalize  recent experiments (Tielrooij et al. 2009).  This possibility, however, cannot be  ruled out  completely because it could be associated to 
the presence of heterogeneities, such as those associated with water molecules with a single hydrogen bond to a lipid at low hydration (Volkov et al. 2007).

Following the structural and dynamical characterization of  water near a membrane,
it is possible to classify water as inner-membrane, first hydration layer and outer-membrane  water. The inner-membrane water has a structural role forming bridges between  lipids (Lopez et al. 2004, Pasenkiewicz-Gierula et al. 1997). This water has a dynamics up to two orders of magnitude slower than bulk water
as a consequence of the robustness of water-lipid hydrogen bonds, which are more frequent the lower the hydration of the membrane.

In the first hydration layer  water-water hydrogen bonds have a dynamics one order of magnitude slower than bulk water. This is due to 
 longer hydrogen bonds lifetime  the lower the hydration,  a consequence of the slowdown of hydrogen bond-switching due to the decrease of water density (Calero et al. 2016).
 
 These effects for the inner-membrane water and the first hydration layer are emphasized when the temperature decreases. We show  that water near the membrane  has a glassy-like behavior when $T=288.6$ K. 
 In particular, 
 the rotational correlation function of water  within 3 \AA~from the membrane  is comparable, in terms of correlation time, to that of  
 bulk water at $\approx$ 30 K colder temperature, but with a characteristic stretched exponent, $\beta(T=288.6$~K) = 0.44, that is much smaller than that of the bulk case, $\beta_{\rm bulk}(T=250$~K) = 0.85 (Kumar et al. 2006), suggesting a larger heterogeneity of relaxation modes in the membrane hydration water. 
 
The water slowing down decreases rapidly for water at larger distance from the membrane. However, the dynamics, as well as density, 
 of bulk water are recovered at a short distance, $\simeq 1.2$ nm from the surface of the membrane, the perturbation on the structural properties of liquid water propagates over a much larger distances, at least $\simeq 2.5$ nm (Martelli, Ko, Borallo and Franzese 2018).
 
This  effect is emphasized by the adoption of a  local  structural parameter $S$ that
has a higher sensitivity, with respect to
the correlation functions, to small effects arising from the interface. In particular,    adopting a definition based on  the second water hydration shell,  and considering that water diameter and hydrogen bond length are quite similar,
$S$ is able to capture  structural effects extending over nine water diameters.
It is worthy to mention that short-range order parameters--e.g.,  the tetrahedrality--are unable to show such a results and are recovering bulk values at the same 
 distances where the density and the dynamical properties do.

Finally, water molecules interacting with P in phospholipids are slightly more ordered those interacting with N. Moreover, water molecules interacting with P-groups have closer second neighbors than those interacting with N-heads. These structural observations, together with the dynamical analysis of hydration water could help us to understand the role of water at biomembrane interfaces (Calero and Franzese 2018).
\begin{acknowledgments}
We are thankful to Fabio Leoni  for helpful discussions. We acknowledge the support of Spanish MINECO grant FIS2015-66879-C2-2-P.
F.M. ackowledges the support of the STFC Hartree Centre's Innovation Return on Research programme, funded by the Department for 
Business, Energy \& Industrial Strategy.

\end{acknowledgments}

%\bibliography{Capitol_v4.bbl}

\end{document}